\documentclass[12pt]{article}

\usepackage{amsmath, amssymb, amsthm} 
\usepackage{hyperref}
\usepackage{graphicx}
\usepackage{cite} 
\usepackage{authblk} 
\usepackage{physics}
\usepackage{tikz}
\usepackage{mathdots}
\usepackage{yhmath}
\usepackage{float}
\usepackage{mathrsfs}  
\usepackage{cancel}
\usepackage{color}
\usepackage{siunitx}
\usepackage{array}
\usepackage{multirow}
\usepackage{gensymb}
\usepackage{tabularx}
\usepackage[most]{tcolorbox} 
\usepackage{extarrows}
\usepackage{booktabs}
\hypersetup{
    colorlinks=true,
    linkcolor=blue,
    citecolor=blue,
    urlcolor=blue
}
\usetikzlibrary{fadings,patterns,shadows.blur,shapes}

\usepackage[left=2.5cm,right=2.5cm,top=2.5cm,bottom=3cm]{geometry} 
\title{\textbf{Holography vs. Scale Separation}}
\author[1]{Alek Bedroya}
\author[2]{Paul J.~Steinhardt}

\affil[1]{\small Princeton Gravity Initiative, Princeton University, Princeton, NJ 08544, USA}
\affil[2]{\small Department of Physics, Princeton University, Princeton, NJ 08544, USA}

\date{} 

\begin{document}

\maketitle

\vskip 10mm

\begin{abstract}
In this work, we point out a contradiction between holography and scale-separated AdS (i.e. parametrically large mass gap) in string theory, making the standard assumption that the holographic CFT describes the IR degrees of freedom on a brane that decouple from gravity. We show that the CFT can only decouple from gravity if the scalar potential in the dual AdS satisfies a certain criterion. Namely, there must exist a scalar field trajectory that follows the gradient of the scalar potential to the asymptotic region of the scalar field space in which limit $\partial_\phi\ln(V)\partial_\phi\ln(\Lambda_s)\leq2/(d-2)$, where $\Lambda_s$ is the quantum gravity cut-off. This condition, which generically implies lack of scale-separation, is satisfied in the standard examples of AdS/CFT. However, proposed attempts at achieving scale separation, such as DGKT, employ scalar potentials that violate this condition. We therefore conclude that the CFT duals of DGKT vacua cannot exist in string theory. Barring fine-tuning, our conclusions apply to other Ricci-flat flux compactifications including the KKLT scenario which relies on scale separation to obtain a metastable de Sitter uplift.
\end{abstract}

\newpage

\tableofcontents
\section{Introduction}

UV completing a gravitational effective field theory is notoriously more difficult than for an ordinary quantum field theory. A key reason is that sufficiently massive states are black holes, which obey universal properties any UV completion must respect. Although a precise UV completion that can answer, for example, scattering experiments at arbitrarily high energies remains out of reach, string theory has provided compelling evidence that certain theories are UV completable.

Two structural features underpin this success. First, string theory admits weak coupling limits, where the string coupling goes to zero and the theory can be controlled at arbitrarily high energies compared to the Planck scale. Second, string dualities relate these weakly coupled descriptions to each other, ensuring consistency across the full moduli space at finite couplings. Indeed, almost all precise results in string theory exploit these weak coupling regimes.

Such limits exhibit universal behavior captured by the Swampland Distance Conjecture~\cite{Ooguri:2006in}. As the expectation value of a scalar field is taken to infinity, a tower of light states emerges with exponentially decreasing masses, and the quantum gravity cutoff falls accordingly~\cite{vandeHeisteeg:2023uxj}. In Minkowski space this provides a consistent weakly coupled regime, but in curved backgrounds the situation is more subtle. 

In AdS, moving to infinite distance in moduli space drives the cutoff below the curvature scale, invalidating the gravitational description. This is precisely the regime where the dual CFT becomes weakly coupled. Therefore, unlike in Minkowski space, there is no arbitrarily weakly coupled gravitational description of AdS. In the absence of infinite-distance limits where the gravitational description remains valid, there are two natural ways to generalize the Distance Conjecture to AdS spaces. One is to rely on the dual CFT~\cite{Perlmutter:2020buo}; the other is to take limits in the space of AdS solutions, rather than in the moduli space of a single AdS. Since the moduli parametrize the space of solutions, the latter approach is natural. Following this logic, the authors of~\cite{Lust:2019zwm} conjectured that there exist constants \( A \) and \( \alpha \), depending only on the spacetime dimension, such that a tower of states appears with mass spacing $m$ such that
\begin{align}
  A m^\alpha < |\Lambda_{\rm AdS}| \,,
\end{align}
in Planck units. A stronger form of the conjecture sets \( \alpha = 2 \), asserting the existence of a tower of light states satisfying
\begin{align}
  A m^2 \lesssim |\Lambda_{\text{AdS}}| \,.
\end{align}
This \emph{no-scale-separation conjecture} states that the AdS scale cannot be parametrically separated from the KK scale~\footnote{We comment on the difference between the version of the AdS Distance Conjecture used here and the strong AdS Distance Conjecture formulated in~\cite{Lust:2019zwm} in Section~\ref{ADSSS}.}

We approach this question holographically. Constructions with curved internal geometries naturally produce \( |\Lambda_{\rm AdS}| \sim m_{\rm KK}^2 \), since the curvature of the internal space contributes terms of order \( m_{\rm KK}^2 \) to the potential that help stabilize the volume modulus. However, proposed scale-separated vacua such as DGKT~\cite{DeWolfe:2005uu} and KKLT~\cite{Kachru:2003aw} instead rely on Ricci-flat compactifications with fluxes. In these cases the scalar potential is not proportional to \( m^2 \), raising the possibility of scale separation. It is important to settle the validity of proposed scale–separated AdS vacua, or to understand precisely what goes wrong in those constructions. For example, KKLT AdS vacua serve as a stepping stone to KKLT de Sitter constructions, and the question of their validity carries important implications for the string landscape and its role in cosmology. By contrast, DGKT offers a more controlled scenario: unlike KKLT, which cannot be parametrically controlled (since the $\alpha'$ corrections cannot be made arbitrarily small), DGKT admits families of AdS solutions labeled by \( N \) that, in principle, allow for arbitrarily small Ultimately, the question of whether an AdS background is UV completable in quantum gravity is decided not by the gravitational description, but by the existence of its dual CFT. This is because the gravitational description can never become arbitrarily weakly coupled: as soon as one takes any coupling to zero, the appropriate description switches to that of the CFT. Therefore, while scale separation in terms of string coupling and internal volume may suggest parametric control, true UV completeness of an AdS background requires identifying the dual CFT.

Our key observation is that the brane supporting the dual CFT can be described as a scalar field solution of the AdS effective action, where the scalar field runs to infinity. This perspective allows us to test whether problematic asymptotic behavior of the potential—specifically, behavior that develops scale separation in the interior—obstructs the existence of a decoupling limit. We show that for a brane theory to decouple from bulk gravity, the scalar potential at infinity must satisfy \( \partial_\phi\ln(V)\partial_\phi\ln(\Lambda_s)  \leq 2/(d-2) \) along the direction of steepest ascent of $V$, where $\Lambda_s$ is the quantum gravity cut-off. Otherwise, the worldvolume CFT cannot be realized as a brane theory in string theory. In particular, if DGKT AdS exists, then string theory would fail to contain its superconformal field theory (SCFT) dual. This condition is relevant for scalar separation, because it implies that $m^2\lesssim V$ along the direction of steepest descent of $ \Lambda_s$ in the moduli space. 

This idea, that varying scalar backgrounds encode essential consistency conditions beyond isolated critical points, has proven fruitful in constraining positive potentials as well~\cite{Bedroya:2024zta,Bedroya:2022tbh}. 

Finally, we extend our methods to de Sitter. Interpreted in the context of dS/CFT~\cite{Strominger:2001pn}, our analysis shows that no Euclidean CFT dual to de Sitter can be realized on a spacelike brane in string theory. This conclusion is qualitatively different from the AdS case: while only scale-separated AdS vacua fail the decoupling test,  no dual can exist for de Sitter.

The organization of the paper is as follows. In Section~\ref{sec2}, we review what is known about infinite-distance limits in the string landscape and introduce the motivation for the absence of scale separation in AdS. There, we also provide a novel formulation of the no scale-separation condition that can be stated sharply within the gravitational description, without relying on the CFT. This formulation is based on the asymptotic behavior of the scalar potential, and we refer to it as the \textit{Asymptotic No-Scale-Separation Condition}. In Section~\ref{Ext}, we turn to extremal black branes in supergravity and analyze their near-horizon geometries. Section~\ref{dec} identifies a crucial decoupling condition for holography. Namely, suppose the holographic dual is realized as a worldvolume theory on a brane. In order to decouple this theory from bulk gravity at low energies, the energy of modes in the surrounding spacetime must remain bounded above in units of the quantum gravity cutoff as one moves away from the brane. In Section~\ref{DC}, we develop a general method to construct black brane geometries from the AdS effective action as scalar field solutions with two spatial infinities, one corresponding to the AdS throat and the other to the asymptotic scalar-field runaway. Building on this, Section~\ref{test} demonstrates that the decoupling condition fails unless the scalar potential satisfies \( \partial_\phi\ln(V)\partial_\phi\ln(\Lambda_s)  \geq 2/(d-2) \) along the direction of steepest ascent of $V$. This matches the asymptotic condition identified in Section~\ref{sec2}, which provides an obstruction to scale separation. In particular, the DGKT construction fails the decoupling test, implying that its purported CFT dual cannot be realized on a brane. Finally, in Section~\ref{dSR}, we extend our analysis to dS/CFT and uncover a fundamental obstruction: no Euclidean CFT dual of de Sitter space can be realized as the worldvolume theory of a brane. We summarize our conclusions in Section~\ref{Conc}.

\section{AdS Scale separation}\label{sec2}

In this section we briefly review the topic of scale separation in AdS, starting with its motivation. The AdS spaces constructed in string theory, with the exception of a few proposals, have compact internal dimensions whose size is of the order of the AdS length scale. This, in particular, implies that AdS spaces come with a tower of light Kaluza-Klein states whose masses scale as $|\Lambda_{\rm AdS}|^{1/2}$, where $\Lambda_{\rm AdS}$ is the AdS cosmological constant. The strong version of the AdS Distance Conjecture quantifies this observation. To review it, let us first recall the Distance Conjecture, which forms its basis.

\subsection{Distance conjecture}

The massless scalar fields in a theory parametrize a subspace of solutions, known as the moduli space $\mathcal{M}$, which preserves the same asymptotic spacetime structure. Depending on the sign of the scalar potential, these moduli parametrize solutions that are Minkowski, dS, or AdS. For example, when the scalar potential vanishes, changing the boundary value of exactly massless scalar fields yields different Minkowski vacua. The moduli space is equipped with a canonical metric induced by the kinetic terms of the scalars:
\[
\mathcal{L}_{\rm Kinetic} = -\frac{1}{2} G_{IJ}(\Phi) \, \partial_\mu \Phi^I \partial^\mu \Phi^J\,,
\]
where $\Phi^I$ is a local parameterization of the massless scalars, and $G_{IJ}$ is the metric on the moduli space. This allows one to define geodesic distance on $\mathcal{M}$.

The \textit{Distance Conjecture }~\cite{Ooguri:2006in} states that as one moves to infinite distance in moduli space, a tower of states becomes exponentially light. More precisely, consider a geodesic $\gamma(t)$ with infinite affine parameter $t \in \mathbb{R}^+$ such that for any $p \in \mathcal{M}$,
\[
\lim_{t \rightarrow \infty} \text{dist}_G(p, \gamma(t)) \rightarrow \infty\,.
\]
Then there exists a tower of particles whose masses scale as
\[
\text{mass} \sim \exp\left(-c \, \text{dist}_G(p, \gamma(t))\right)\,.
\]

An important refinement of the Distance Conjecture is the \textit{Emergent String Conjecture}~\cite{Lee:2019wij}, which postulates that the light tower must fall into one of the following two categories:

\begin{itemize}
    \item \textbf{Decompactification:} The theory decompactifies to a higher-dimensional theory as the volume of internal dimensions grows along $\gamma(t)$. The associated KK tower becomes exponentially light. In the absence of warping, the coefficient $c$ for decompactification from $d$ to $D>d$ dimensions is
    \[
    c_{\rm KK} = \sqrt{\frac{D-2}{(D-d)(d-2)}}\,.
    \]
    
    \item \textbf{String limit:} The theory flows to a perturbative string theory as the dilaton $\varphi$ controlling the string coupling $g_s=\exp(-\varphi\sqrt{d-2}/2)$ diverges. In this limit, the tower of string excitations, with mass proportional to the string scale, becomes light. When only the dilaton varies, the coefficient is
    \[
    c_{\rm string} = \sqrt{\frac{1}{d-2}}\,.
    \]
\end{itemize}

This conjecture is non-trivial even within string theory. For instance, in 4d $\mathcal{N}=2$ theories arising from compactifying type II strings on Calabi-Yau threefolds, it is a non-trivial fact that every infinite-distance limit of the vector multiplet moduli space is of this form~\cite{Friedrich:2025gvs}. In~\cite{Bedroya:2024ubj}, the authors provided a bottom-up argument, based on gravitational amplitudes and black hole thermodynamics, explaining why if the lightest tower is not a KK tower, it must instead exhibit an exponential density of states, characteristic of a string tower. However, from the bottom-up perspective, the precise emergence of the tower at infinite distance in moduli space, as well as the detailed dependence of the tower's mass scale on the scalar fields, remains unclear (see also~\cite{Basile:2023blg,Herraez:2024kux} for alternative approaches to motivating the Emergent String Conjecture).

Furthermore, if we take the Emergent String Conjecture seriously, we find that the coefficient $c$ must obey the inequality
\begin{align}
 c \geq \frac{1}{\sqrt{d-2}}\,,   
\end{align}
which is saturated only for the string limit. This inequality is referred to as the \textit{Sharpened Distance Conjecture}~\cite{Etheredge:2022opl}, and it is preserved under dimensional reduction.

\subsection{AdS distance conjecture}\label{AdSdist}

For Minkowski vacua, we can think of the moduli space as a space of Minkowski solutions and interpret the Distance Conjecture in terms of what happens as one moves to an infinite distance limit in this space of solutions.

\begin{figure}[H]
    \centering

\tikzset{every picture/.style={line width=0.75pt}} 

\begin{tikzpicture}[x=0.75pt,y=0.75pt,yscale=-1,xscale=1]

\draw    (93.97,22.97) .. controls (370.06,241.42) and (448.3,128.89) .. (628.26,19) ;
\draw    (523.19,391.03) .. controls (392.42,180.52) and (479.6,147.42) .. (630.5,29.59) ;
\draw    (512.02,395) .. controls (395.77,220.24) and (282.87,244.07) .. (144.27,395) ;
\draw    (89.5,36.21) .. controls (232.57,151.39) and (279.52,236.13) .. (140.92,387.06) ;
\draw  [fill={rgb, 255:red, 155; green, 155; blue, 155 }  ,fill opacity=0.47 ] (293.84,203.37) .. controls (293.84,183.14) and (320.41,166.75) .. (353.17,166.75) .. controls (385.94,166.75) and (412.5,183.14) .. (412.5,203.37) .. controls (412.5,223.6) and (385.94,240) .. (353.17,240) .. controls (320.41,240) and (293.84,223.6) .. (293.84,203.37) -- cycle ;
\draw    (239.5,140.5) -- (156.5,140.5) ;
\draw [shift={(154.5,140.5)}, rotate = 360] [color={rgb, 255:red, 0; green, 0; blue, 0 }  ][line width=0.75]    (10.93,-3.29) .. controls (6.95,-1.4) and (3.31,-0.3) .. (0,0) .. controls (3.31,0.3) and (6.95,1.4) .. (10.93,3.29)   ;
\draw    (348.17,228) -- (202.5,228.49) ;
\draw [shift={(200.5,228.5)}, rotate = 359.81] [color={rgb, 255:red, 0; green, 0; blue, 0 }  ][line width=0.75]    (10.93,-3.29) .. controls (6.95,-1.4) and (3.31,-0.3) .. (0,0) .. controls (3.31,0.3) and (6.95,1.4) .. (10.93,3.29)   ;

\draw (24,119) node [anchor=north west][inner sep=0.75pt]   [align=left] {\begin{minipage}[lt]{85.5pt}\setlength\topsep{0pt}
\begin{center}
CFT description\\ is weakly coupled
\end{center}

\end{minipage}};
\draw (2,217) node [anchor=north west][inner sep=0.75pt]   [align=left] {\begin{minipage}[lt]{137.09pt}\setlength\topsep{0pt}
\begin{center}
AdS Gravitational description \\is weakly coupled
\end{center}

\end{minipage}};
\draw (305,197.4) node [anchor=north west][inner sep=0.75pt]  [font=\small]  {$| \Lambda _{\text{AdS}} |^{1/2} < \Lambda _{s}$};
\draw (215,151.4) node [anchor=north west][inner sep=0.75pt]  [font=\small]  {$| \Lambda _{\text{AdS}} |^{1/2}  >\Lambda _{s}$};

\end{tikzpicture}
    \caption{The above figure illustrates the massless moduli space of an AdS vacuum, or equivalently, in the CFT language, the conformal manifold of its dual CFT. Except for a finite region where the mass scale $|\Lambda_{\rm AdS}|^{1/2}$ associated with the AdS curvature remains below the quantum gravity cutoff $\Lambda_s$ the AdS description becomes strongly coupled, while the CFT description provides a weakly coupled framework. The quantum gravity cutoff is set by the string scale or the higher-dimensional Planck scale.}
    \label{AdS/CFTmoduli}
\end{figure}
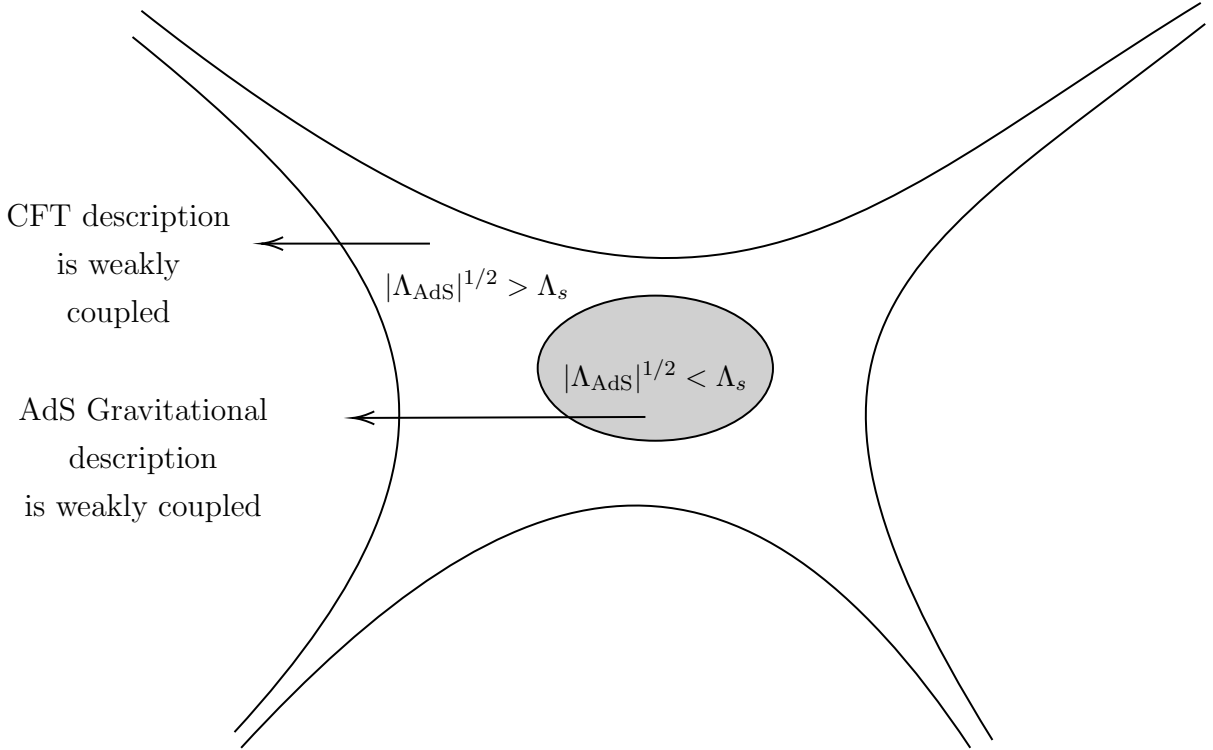

Now let us apply similar reasoning to AdS space. Here, we encounter a key challenge: for any AdS vacuum with massless moduli, the gravitational effective action is only valid over a limited range in moduli space. For instance, in the case of type IIB on $\text{AdS}_5 \times S^5$, if the string coupling becomes too small, the curvature exceeds the string scale, causing a breakdown of the gravitational action. This regime is precisely where the theory is better described by its dual CFT. Conversely, if the string coupling becomes too large, one must use the S-dual type IIB description, which again breaks down due to weak coupling. In other words, there exists only a finite region of the continuous moduli space where the supergravity approximation remains valid (Figure~\ref{AdS/CFTmoduli}). Ultimately, it is only the CFT description that can ever be parametrically weakly coupled, and the only way to establish the UV completeness of a given AdS background is through the construction of its dual CFT.

This limitation is intuitive in light of the Emergent String Conjecture. If we move to an infinite-distance limit in moduli space, the theory either decompactifies or becomes a weakly coupled string theory. In either case, the quantum gravity cutoff—i.e., the energy scale where higher-derivative corrections become important—becomes small in Planck units~\cite{vandeHeisteeg:2022btw}. This cutoff is often referred to as the species scale denoted with $\Lambda_s$~\cite{Dvali:2007hz,Dvali:2007wp,Dvali:2010vm}, due to the qualitative argument that an infinite number of species leads to strong gravitational interactions at lower energies. Let us illustrate this for both the decompactification and string limits:

\begin{itemize}
    \item \textbf{Decompactification:} Higher-derivative corrections in the higher-dimensional theory, suppressed by the higher-dimensional Planck scale, translate into corrections in the lower-dimensional theory. If the curvature exceeds the higher-dimensional Planck scale, the gravitational EFT breaks down. Therefore, the species scale is proportional to the higher-dimensional Planck scale. When the compactification radius is large, the higher-dimensional Planck scale can be much smaller than the lower-dimensional Planck scale.
    
    \item \textbf{String limit:} In this case, $\alpha'$ corrections (stringy corrections) become unsuppressed once the curvature exceeds the string scale, again leading to a breakdown of the gravitational EFT. Therefore, the species scale is set by the string mass in these limits, which can be much smaller than the Planck scale. 
\end{itemize}

Since we understand how the species scale depends on the moduli in both of these cases, we can use the Emergent String Conjecture to infer general behavior in infinite-distance limits. As shown in~\cite{vandeHeisteeg:2023uxj}, the following bounds on the species scale emerge in the infinite distance limits:

\begin{itemize}
    \item Decompactification: $\left| \partial_\phi \Lambda_s / \Lambda_s \right| = \sqrt{\frac{D - d}{(d - 2)(D - 2)}}$
    \item String limit: $\left| \partial_\phi \Lambda_s / \Lambda_s \right| = \frac{1}{\sqrt{d - 2}}$
\end{itemize}

Hence, in the infinite-distance limit, the general inequality is:
\begin{align}\label{ESCA}
\frac{1}{\sqrt{(d-1)(d-2)}} \leq \left| \frac{\partial_\phi \Lambda_s}{\Lambda_s} \right| \leq \frac{1}{\sqrt{d - 2}} \,,
\end{align}
where the lower bound is saturated in decompactification of one compact dimension and the upper bound is saturated in string limits.

If the quantum gravity cutoff $\Lambda_s$ drops below the AdS scale $\sqrt{\Lambda_{\rm AdS}}$, the gravitational action becomes unreliable. But this always occurs in massless directions of moduli space, since the scalar potential—and therefore $\Lambda_{\rm AdS}$—remains fixed, while $\Lambda_s$ decreases. Thus, it is not possible to move to an infinite-distance limit in the space of AdS solutions while maintaining a trustworthy gravitational description; we inevitably transition to a CFT description.

Alternatively, we may consider families of AdS vacua and move between them, but defining a robust notion of distance in this space is subtle. Some proposals have been made in~\cite{Basile:2022zee,Basile:2023rvm,Li:2023gtt,Palti:2024voy,Debusschere:2024rmi,Mohseni:2024njl,Palti:2025ydz}, but a simpler approach is to state the analogue of the Distance Conjecture in terms of the cosmological constant, since in weak-coupling regimes the scalar potential often scales exponentially with the moduli~\cite{Dine:1985he}. Motivated by this, the \textit{AdS Distance Conjecture}~\cite{Lust:2019zwm} replaces the notion of distance with a power-law dependence on $\Lambda_{\rm AdS}$ and states:

There exist constants \( A \) and \( \alpha \), depending only on the spacetime dimension, such that a tower of states appears with mass spacing
\begin{align}
  A m^\alpha <  |\Lambda_{\rm AdS}|\,,  
\end{align}
in Planck units.

It is crucial to emphasize that any AdS space is only well-defined over a finite region of moduli space. The massive moduli are stabilized in the AdS solution, and the massless moduli are bounded because the supergravity description fails outside a limited range. Thus, any statement about AdS geometries pertains to the interior of moduli space. Alternatively, in the CFT dual, which is well-defined for all moduli, these massless bulk moduli correspond to moving in the conformal manifold. This dual description recasts gravitational conjectures into non-trivial statements about CFTs~\cite{Perlmutter:2020buo}. However, since gravity is no longer trustable in that regime, the numerical coefficients in these conjectures can differ substantially.

\subsection{AdS scale separation}\label{ADSSS}

A strong version of the AdS Distance Conjecture sets the exponent $\alpha = 2$, stating that there exists a tower of light states whose masses satisfy
\begin{equation}\label{SADS}
\fbox{\(\;\; \textbf{No-Scale-Separation Conjecture: } 
      Am^2 \;\lesssim\; |\Lambda_{\text{AdS}}| \;\;\)}
\end{equation}

The original strong AdS distance conjecture states that there exists a tower of states whose masses satisfy $m^2 \sim |\Lambda_{\text{AdS}}|$ although this tower is not necessarily the lightest one. For instance, in three-dimensional constructions such as $AdS_3 \times S^3 \times T^4$ in type IIB~\cite{Maldacena:1998bw}, the KK tower from the $T^4$ can be parametrically lighter than $|\Lambda_{\rm AdS}|^{1/2}$ while the KK tower from $S^3$ has a mass scale of $|\Lambda_{\rm AdS}|^{1/2}$. The strong AdS distance conjecture clearly implies the no-scale separation conjecture, but it is stronger.
It was conjectured in~\cite{Perlmutter:2020buo} that in $AdS_d$ with $d>3$, the KK tower cannot be parametrically lighter than $|\Lambda_{\rm AdS}|^{1/2}$. In other words, in $d>3$, the lightest KK tower cannot be decoupled from the AdS scale. 

we adopt the refined version~\eqref{SADS}, which asserts that the above scaling must hold for the lightest tower of states. We refer to this statement as the \textit{no-scale-separation conjecture}. This formulation permits the KK scale to be arbitrarily lighter than the AdS scale, but crucially it forbids the opposite: the KK scale cannot be parametrically heavier than $|\Lambda_{\rm AdS}|^{1/2}$. In other words, when we refer to \emph{no-scale separation}, we mean that arbitrarily heavy KK scales relative to the AdS scale are forbidden, while lighter towers are allowed. 

Importantly, the statement~\eqref{SADS} can hold throughout the entire moduli space of AdS, even in regimes where the gravitational description is strongly coupled and the CFT description is weakly coupled. In such corners, it is typically the tower of string excitations that becomes arbitrarily lighter than the AdS scale. Likewise, in higher-spin theories, the lightest tower can be parametrically lighter than $|\Lambda_{\rm AdS}|^{1/2}$, but never heavier. Vasiliev-type higher spin theories~\cite{Fradkin:1987ks} which can be holographic~\cite{Klebanov:2002ja} are also examples of such theories~\cite{Basile:2022sda}.

The central question is whether one can obtain an AdS vacuum whose curvature scale is parametrically smaller than the energy scale associated with extra physics, such as KK modes or string excitations. Scale-separated AdS spaces are particularly important for scenarios like KKLT \cite{Kachru:2003aw}, which begin with such vacua as a foundational step.\footnote{For an alternative approach, see~\cite{Bena:2022cwb}.}

Let us illustrate this with a basic example: $\text{AdS}_5 \times S^5$. Compactifying 10D type IIB string theory on $S^5$ with $N$ units of self-dual $F_5$ flux gives rise to a 5D Einstein-frame effective theory with a scalar potential composed of two main contributions:
\begin{itemize}
    \item \textbf{Internal curvature:} The positive curvature of $S^5$ contributes negatively to the scalar potential. Since this curvature scales as the inverse square of the radius, it behaves as $m_{\rm KK}^2$, where $m_{\rm KK}$ is the KK mass scale. Hence, this contribution scales like $m_{\rm KK}^2 m_{\text{pl},d}^{d-2}$.
    
    \item \textbf{Fluxes:} Flux quantization in the higher-dimensional theory gives $\int_{S^5} \star F = N$, implying that $|F|^2$ scales inversely with the volume of $S^5$. Therefore, the flux term contributes $\propto N^2 m_{\rm KK}^5$ to the potential.
\end{itemize}

Thus, the effective scalar potential takes the form
\begin{align}
V(\phi) = c_{\text{flux}} N^2 m_{\rm KK}^5 - c_{\text{curvature}} m_{\text{pl},5}^3m_{\rm KK}^2\,,
\end{align}
where $m_{\rm KK}$ depends on the canonically normalized volume modulus as
\begin{align}
m_{\rm KK} = m_{\text{pl},5} \exp\left(-\sqrt{\frac{8}{15}}\phi\right)\,.
\end{align}
The coefficient $c_{\text{curvature}}$ is independent of $N$, and is defined as
\begin{align}
c_{\text{curvature}} = \frac{\mathcal{R}_{S^5}}{m_{\rm KK}^2}\,,
\end{align}
where $\mathcal{R}_{S^5}$ is the scalar curvature of the 5-sphere. Minimizing $V$ with respect to $m_{\rm KK}$ gives
\begin{align}
5 c_{\text{flux}} N^2 m_{\rm KK}^4 = 2 c_{\text{curvature}} m_{\text{pl},5}^3m_{\rm KK}
\quad \Rightarrow \quad
V_{\min} = -\frac{3}{5} c_{\text{curvature}}m_{\text{pl},5}^3 m_{\rm KK}^2\,.
\end{align}

Because the two terms balance at the minimum, the cosmological constant scales as $m_{\rm KK}^2$. The exact prefactor depends on the exponents in the potential, but this behavior is generic: in any scalar potential where internal curvature contributes, this term dominates in the large-volume limit, and other effects (e.g., fluxes or non-perturbative contributions) must compete near the minimum. As a result, the AdS scale is generically tied to the KK scale, making scale separation challenging to achieve in controlled string compactifications.

\subsection{Attempts at Scale Separation}\label{attempts}

There are two strategies for constructing scale-separated AdS vacua. One is to find a family of AdS vacua labeled by $n$ where $\Lambda_{\rm AdS} = A_n m_{\rm KK}^{2}$ with $\lim_{n\rightarrow\infty}A_n \to 0$. The other is to find vacua such that $\Lambda_{\rm AdS} \propto m_{\rm KK}^{\alpha}$ with $\alpha > 2$. From the previous subsection, we saw that the asymptotic behavior of the scalar potential dictates how the cosmological constant behaves. In fact, both strategies attempt to engineer scale-separated AdS by modifying this behavior.

To realize $\Lambda_{\rm AdS} = A_n m_{\rm KK}^{2}$ with $A_n \to 0$, one can retain the curvature term in the scalar potential but suppress its coefficient, which is given by the ratio $\mathcal{R}_{\rm internal}/m_{\rm KK}^2$. Since we are considering infinite distance limits where curvature dominates, the internal space must be an Einstein manifold (possibly with singularities). The mathematical question then becomes: can there exist a family of Einstein manifolds for which $ \mathcal{R}_{\rm internal}/m_{\rm KK}^2\to 0$?

This question was studied in~\cite{Collins:2022nux}. While the internal volume can be made small (e.g., via orbifolds of spheres), the KK scale is controlled by the diameter $D$, which bounds $m_{\rm KK}^{-1}$ from both above and below. The authors of~\cite{Collins:2022nux} showed that for orbifolds of spheres and Calabi-Yau singularities, this limit cannot be achieved. However, the general mathematical question remains open.

A speculative proposal involves compactifying M-theory on a family of weak $G_2$ manifolds (a subclass of Einstein manifolds) with $D^2 \mathcal{R}_{\rm internal} \to 0$~\cite{Cribiori:2021djm}. If such manifolds exist, supersymmetric Freund–Rubin compactifications~\cite{Freund:1980xh} would realize scale separation. In this work, we assume the conjecture of~\cite{Collins:2022nux} is correct, and such manifolds do not exist, implying no geometric realization of AdS scale separation via this route. However, settling this question is of fundamental importance, as it relates to the most reliable and controllable method for achieving scale separation (see~\cite{VanHemelryck:2024bas} for recent work in this direction).

A more exotic strategy is to change the asymptotic behavior of the scalar potential such that it scales as $m_{\rm KK}^{\alpha}$ with $\alpha > 2$ in some infinite direction in moduli space. Then, $\Lambda_{\rm AdS} \sim m_{\rm KK}^{\alpha}$, leading to parametric scale separation. This is the route pursued in constructions like KKLT~\cite{Kachru:2003aw} and DGKT~\cite{DeWolfe:2005uu}. Both approaches try to do this by considering internal geometries that are Ricci flat to first order. In DGKT, unlike KKLT, the volume of the internal volume can be made parametrically large. Moreover, the string coupling can be made parametrically small. Due to this parametric control, DGKT is the most robust proposal to achieve scale separation which is why we focus on it here.

DGKT is a compactification of massive type IIA string theory on an orientifold threefold. We consider one of the most studied DGKT setups: a compactification on $T^6/\mathbb{Z}_3 \times \mathbb{Z}_3$ orientifold~\cite{DeWolfe:2005uu}. The resulting scalar potential is given by
\begin{align} \label{DGKTV}
V = M_{\rm pl}^4\, s^{-3} \left[
  \frac{A_{F_4}}{us} +
  \frac{A_{F_0} u^3}{s} +
  \frac{A_{H_3} s}{u^3} -
  A_{O_6}
\right],
\end{align}
and the moduli $s$ and $u$ are defined as
\begin{align}
    s &= g_s^{-1} \left( \frac{l_{\rm KK}}{l_s} \right)^3, &
    u &= \left( \frac{l_{\rm KK}}{l_s} \right)^2.
\end{align}
The terms in \eqref{DGKTV} are respectfully sourced by $F_4$ flux, Romans mass, $H_3$ flux and Orientifolds. The six-torus $T^6$ is written as the product of three 2-tori, each with complex structure $\tau = e^{2\pi i / 6}$. Let $z_1, z_2, z_3$ be the complex coordinates parametrizing each torus, with identifications $z_i \sim z_i + 1 \sim z_i + \tau$. The $\mathbb{Z}_3 \times \mathbb{Z}_3$ is generated by
\begin{align}
    (z_1, z_2, z_3) &\rightarrow \left(\tau^2 z_1 + \frac{1+\tau}{3},\, \tau^4 z_2 + \frac{1+\tau}{3},\, z_3 + \frac{1+\tau}{3} \right), \\
    (z_1, z_2, z_3) &\rightarrow \tau^2 (z_1, z_2, z_3)\,.
\end{align}
The quotient $T^6/(\mathbb{Z}_3 \times \mathbb{Z}_3)$ has nine orbifold fixed points. Next, we impose an orientifold projection with the involution $z_i \rightarrow \bar{z}_i$ which introduces space-filling O6-planes in the four non-compact dimensions. 

To cancel the F$_2$ tadpole sourced by the O6-planes, we turn on $H_3$ flux and Romans mass $F_0$, satisfying
\begin{align}
    \int dF_2 = \int \delta_{\text{O6}} + m\,H_3 = 0\,.
\end{align}
This condition implies that the product of the Romans mass and the $H_3$ flux must cancel the O6-plane contribution. Consequently, in the scalar potential~\eqref{DGKTV}, the orientifold contribution is fixed by
\begin{align}
    A_{O_6}^2 = 16 A_{F_0} A_{H_3}\,.
\end{align}

Finally, we include arbitrary $F_4$ fluxes along orientifold-even 4-cycles, leading to
\begin{align}
    N = \int F_4 \,, \quad \Rightarrow \quad A_{F_4} = N^2\,.
\end{align}

The third term in \eqref{DGKTV} is never dominant, however, the second and fourth terms compete for large internal volumes and stabilize the string coupling. Balancing the second and fourth terms leads to
\begin{align}
s &\sim \left( \frac{l_{\rm KK}}{l_{\rm pl}} \right)^{6/7}, &
u &\sim \left( \frac{l_{\rm KK}}{l_{\rm pl}} \right)^{2/7}, \\
g_s &\sim \left( \frac{l_{\rm KK}}{l_{\rm pl}} \right)^{-3/7}, &
\frac{l_s}{l_{\rm pl}} &\sim \left( \frac{l_{\rm KK}}{l_{\rm pl}} \right)^{6/7}, &
\frac{l_s}{l_{\rm KK}} &\sim \left( \frac{l_{\rm KK}}{l_{\rm pl}} \right)^{-1/7}.
\end{align}

Thus, the potential becomes
\begin{align}
V \simeq M_{\rm pl}^4 \left[
A N^2 \left( \frac{m_{\rm KK}}{M_{\rm pl}} \right)^{26/7}
- B \left( \frac{m_{\rm KK}}{M_{\rm pl}} \right)^{18/7}
\right],
\end{align}
where $A$ and $B$ are constants. This asymptotic behavior differs from $m_{\rm KK}^2$, enabling parametric scale separation. 

\subsection{Asymptotic No-Scale-Separation}\label{ANSS}

The mechanism described above for modifying the exponent of $m_{\rm KK}$ is precisely the type of scale separation we aim to test. Achieving such separation requires the scalar potential to fall off faster than $m^2$ in the asymptotic regions of field space. As shown in~\cite{Rudelius:2022gbz,Andriot:2020lea,Castellano:2023jjt}, in any infinite-distance limit one has the identity
\begin{align}
    \frac{\nabla m}{m}\cdot \frac{\nabla \Lambda_s}{\Lambda_s}
    = \frac{1}{d-2}\,,
\end{align}
where $\Lambda_s$ is the quantum-gravity cutoff. Thus, naively, if $m^2$ decays faster than $V$ along the direction of $\nabla \Lambda_s$, one would expect
\begin{align}\label{min}
    \frac{\nabla V}{V} \cdot \frac{\nabla \Lambda_s}{\Lambda_s}
    \leq \frac{2}{d-2}\,.
\end{align}

Motivated by this, we propose the following criterion for the absence of scale separation:

\begin{center}
\setlength{\fboxsep}{8pt}
\setlength{\fboxrule}{1pt}
\fbox{
  \begin{minipage}{0.9\textwidth}
  \vspace{4pt}
  \textbf{Asymptotic No-Scale-Separation Condition (ANSS):}\\[4pt]
  \emph{For any scalar potential with an AdS critical point, there exists an infinite-distance limit along $\nabla V$ in which the scalar potential remains negative and}
  \[
    \frac{\nabla V}{V}\cdot\frac{\nabla \Lambda_s}{\Lambda_s}
    \leq \frac{2}{d-2}\,,
  \]
  \emph{where $\Lambda_s$ is the quantum-gravity cutoff.}
  \vspace{4pt}
  \end{minipage}
}
\end{center}

\noindent
A clarifying remark: by an \emph{infinite-distance limit}, we mean a direction in field space along $\nabla V$ for which $V\to 0$. This explicitly excludes limits such as shrinking the $S^5$ in $\text{AdS}_5\times S^5$, where $V\to +\infty$ and the effective field theory breaks down. This technical point will become important in Section~\ref{test}.

We will show that this condition is intimately tied to the existence of a standard holographic dual. In particular, it is powerful enough to obstruct all known proposals for parametrically scale-separated \emph{stable} AdS vacua.\footnote{The AdS background must satisfy the Breitenlohner--Freedman bound $m^2 l_{\rm AdS}^2 \geq -\tfrac{(d-1)^2}{4}$~\cite{Breitenlohner:1982bm}. Tachyonic modes above the BF bound correspond to relevant operators in the dual CFT.} This is why we refer to it as a \emph{condition} rather than a conjecture. Nevertheless, the more direct statement below is also quite natural:

\begin{center}
\setlength{\fboxsep}{8pt}
\setlength{\fboxrule}{1pt}
\fbox{
  \begin{minipage}{0.9\textwidth}
  \vspace{4pt}
  \textbf{Strong Asymptotic No-Scale-Separation Conjecture:}\\[2pt]
  \emph{For any scalar potential with an AdS critical point, there exists an infinite-distance limit along $\nabla V$  in which $V<0$ and}
  \[
     \lim_{\phi\rightarrow\infty}\frac{m^2}{|V|} < \infty\,,
  \]
  \emph{where $m$ is the mass scale of the lightest tower.}
  \vspace{4pt}
  \end{minipage}
}
\end{center}

\noindent
There is an important limitation in interpreting the strong AdS distance conjecture~\eqref{SADS} in terms of the KK tower. The gravitational description is valid only in a finite region of moduli space; in this region the KK tower is typically unstable, and its mass spectrum is blurred by interactions. If one instead approaches a weakly coupled limit in AdS\(_d\) with \( d>3 \), the lightest tower is \emph{never} a KK tower, but rather a higher-spin tower associated with string excitations~\cite{Perlmutter:2020buo}. This is precisely the content of the \emph{CFT distance conjecture}. Thus, one encounters a basic puzzle: no trajectory in AdS moduli space allows the KK tower to remain the lightest tower, and the gravitational description cannot be taken to arbitrarily weak coupling. Any purely gravitational statement is therefore approximate.

In contrast, the Asymptotic No-Scale-Separation Condition avoids this difficulty entirely. It is formulated directly in the infinite-distance limit where the KK tower \emph{is} the relevant tower. One might wonder whether this limit inevitably leads to an EFT breakdown, as happens when moving along moduli directions of AdS. However, as we will demonstrate in Section~\ref{DC}, the scalar potential is \emph{not} flat in this direction, and this allows us to construct smooth solutions in which the scalar field runs to infinity without encountering an EFT breakdown.

This motivates the approximate claim that in any AdS/CFT setup that admits an Einstein-gravity regime, there must exist a KK tower whose mass satisfies
\begin{align}
   A\, m \;\leq\; |\Lambda_{\rm AdS}|^{1/2}
\end{align}
for some positive constant \( A \). If a tower is parametrically lighter than \( |\Lambda_{\rm AdS}|^{1/2} \), it cannot be a string tower, as $\alpha'$ corrections would invalidate the Einstein-gravity approximation. Since this argument within gravity is only approximate, we instead rely on the ANSS condition, which is sharp in the gravitational description and—absent fine tuning—implies the inequality above. Indeed, if $m^2 \lesssim |V|$ in the direction of $\nabla \Lambda_s$, as required by ANSS, then when flowing back from the asymptotic region toward the AdS critical point, other contributions to $V$ must compete with (or dominate) the asymptotic behavior. Hence, barring fine tuning,
\[
|\Lambda_{\rm AdS}|
   \;\gtrsim\; m^2,
\]
and the AdS vacuum fails to be scale separated.

\medskip

\noindent
For clarity, we now summarize the various refinements of the AdS Distance Conjecture:

\begin{itemize}
    \item \textbf{Strong AdS Distance Conjecture:}  
    There exists a tower of states whose mass satisfies  
    \[
      m \sim |\Lambda_{\rm AdS}|^{1/2},
    \]
    forbidding parametrically separating the AdS scale from this tower.

    \item \textbf{No-Scale-Separation Conjecture:}  
    The \emph{lightest} tower satisfies  
    \[
      m \lesssim |\Lambda_{\rm AdS}|^{1/2},
    \]
    forbidding the lightest tower from becoming parametrically heavier than the AdS scale (while allowing additional lighter towers).

    \item \textbf{Asymptotic No-Scale-Separation Condition (ANSS):}  
    In some infinite-distance limit along $\nabla V$,  
    \[
      \partial_\phi \ln V \,\partial_\phi \ln \Lambda_s 
      \leq \frac{2}{d-2},
    \]
    a sharp, purely gravitational condition that we prove in Section~\ref{test}.

    \item \textbf{Strong Asymptotic No-Scale-Separation Conjecture:}  
    In an infinite-distance limit with along $\nabla V$, $V\rightarrow 0^-$ and,  
    \[
      m^2 \lesssim |V|,
    \]
    the most direct formulation in terms of scale separation.
\end{itemize}

\noindent
We will prove the Asymptotic No-Scale-Separation Condition in Section~\ref{test} by deriving a contradiction with holography.

Finally, note that the existence of an AdS critical point is crucial. For example, in a Scherk--Schwarz compactification of type IIB on a circle with antiperiodic fermions, the potential behaves as $V\sim m_{\rm KK}^9$ but has no global minimum due to a tachyonic winding mode. In such cases the ANSS criterion is inapplicable.\footnote{See~\cite{Aparici:2025kjj} for related instabilities.}  
Indeed, the ANSS condition can be used as a diagnostic: if the asymptotic behavior of $V$ violates it, the theory cannot possess a stable AdS critical point.

\vspace{1em}

\section{Extremal Black Branes}\label{Ext}

In this section, we review the universal properties of extremal black branes in supergravity. We begin with the simplest type of extremal branes, which sit in asymptotically Minkowski spacetimes and have regular horizons. In Subsection~\ref{sing}, we generalize our discussion to extremal branes with singular horizons, and in Subsection~\ref{exotic}, to extremal branes with other asymptotic spacetimes such as AdS.

\subsection{Regular horizons in Minkowski}\label{MinB}

The metric for a general black $p$-brane in asymptotically Minkowski spacetime takes the form 
\begin{align}
    ds^2 \;=\; -\lambda_1(r) \, dt^2 \;+\; \lambda_2(r)\, d\vec{y_p}^2 \;+\; \mu(r)^{-1}\, dr^2 \;+\; R(r)^2\, d\Omega_{D-p-2}^2 \,,
\end{align}
where $\vec{y}_p$ are the spatial coordinates parallel to the brane, $r$ parametrizes the radial distance from the brane, and $d\Omega_{D-p-2}^2$ denotes the metric on the $(D-p-2)$-sphere capturing the angular directions transverse to the brane.
There are three key features that distinguish extremal $p$-branes with regular horizons.

\subsubsection*{Infinite throat}

The horizon lies at infinite proper radial distance. This occurs because extremal branes have coincident horizons, meaning $\mu(r)$ has a double zero. As a result, the proper radial distance
\[
\Delta\rho=\int_{r_{\min}}^{r_{\max}} \mu(r)^{-1/2}\, dr
\]
diverges as $r_{\min}$ approaches the double zero of $\mu(r)$. Thus, in terms of the proper distance coordinate $\rho$, the radial coordinate spans $(-\infty, +\infty)$. Note, however, that the horizon is at infinite proper distance only on constant-time slices. An infalling observer crosses the horizon in finite proper time; the horizon is a null surface and part of the extended geometry.

\subsubsection*{Worldvolume Lorentz symmetry}

Another important property of extremal black branes is the Lorentz symmetry along the directions parallel to the brane. For BPS black branes, this symmetry arises from the preserved part of the supersymmetry algebra. Lorentz invariance implies that coordinates can be rescaled so that the coefficients of $dt^2$ and $d\vec{y}_p^2$ match. Hence, the extremal black brane metric takes the form
\begin{align}\label{metricform1}
    ds^2 \;=\; f(\rho)^2 \big[-dt^2 + d\vec{y}_p^2\big] \;+\; d\rho^2 \;+\; R(\rho)^2\, d\Omega_{D-p-2}^2 \,.
\end{align}

\subsubsection*{AdS near-horizon geometry}

For black branes with regular horizons, the size of the transverse sphere asymptotes to a constant as $\rho \rightarrow -\infty$. To prevent curvature singularities in this limit, the curvature of the $(p+2)$-dimensional Lorentzian space spanned by $\{t, \vec{y}_p, \rho\}$ must remain finite. The relevant components of the Ricci tensor are
\begin{align}
    \mathcal{R}^{\mu}_{\phantom{\mu}\mu} &= -\partial_\rho^2 \ln f - p \left(\partial_\rho \ln f \right)^2\,, \\
    \mathcal{R}^\rho_{\phantom{\rho}\rho} &= -(p+1) \partial_\rho^2 \ln f\,.
\end{align}
These expressions are finite if and only if $f(\rho)$ behaves exponentially:
\[
f(\rho) \sim e^{\rho / L}\,.
\]
Therefore, the near-horizon geometry of extremal black branes with regular horizons takes the form
\begin{align}
    ds^2 \;\simeq\; e^{2\rho/L}\big[-dt^2 + d\vec{y}_p^2\big] + d\rho^2 + R^2\, d\Omega_{D-p-2}^2\,,
\end{align}
which is locally the product geometry $\text{AdS}_{p+2} \times S^{D-p-2}$, with AdS radius $L$. This confirms that the horizon at $\rho = -\infty$ is traversable in finite proper time by an infalling observer since it corresponds to the Poincaré horizon of AdS.

\subsection{Singular horizons}\label{sing}

To obtain extremal black branes with regular horizons, the gauge coupling of the field strength under which the brane is charged must attain a non-zero minimum at finite values of the moduli. For instance, suppose the brane is charged under a set of $(p+1)$-form gauge potentials with kinetic terms of the form
\begin{align}
   \mathcal{L}_F = -\frac{M_{\rm pl}^{D-2}}{4(p+2)!}\,K^{IJ}(\Phi)\,F_I F_J \,,
\end{align}
and carries integer-valued charges $q_I = \int \star F_I$, where the integral is over a sphere that links with the brane. We define the effective charge $q_{\rm eff}$ such that 
\begin{align}
    q_{\rm eff}^2 = \frac{M_{\rm pl}^{D-2}}{2}q_I q_J K^{IJ}(\Phi)\,.
\end{align}
If the brane is charged under different gauge fields, $q_{\rm eff}$ represents the effective gauge coupling of the combined gauge field that couples to the black brane. The moduli of the extremal brane vary over space such that $q_{\rm eff}$ is minimized on the horizon~\cite{Ferrara:1997tw,Gibbons:1996af}. This behavior is known as the \textit{attractor mechanism}. Intuitively, it means that near the black brane horizon, the scalar fields adjust themselves to minimize the effective mass contribution from the core of the black brane. One can heuristically divide the total mass into two contributions: (i) the energy stored in the scalar field profile, and (ii) the energy due to the core of the extremal black brane. The scalar fields evolve to minimize the latter, which is governed by the effective gauge coupling evaluated at the horizon. 

If $2(p+2) = d$, then there can be additional kinetic couplings between the gauge field strengths and their duals. Moreover, the charged black branes can carry \emph{dyonic charges}, because the magnetically charged black branes have the same dimension as the electrically charged ones. Suppose the kinetic mixing between the gauge field strengths and their Hodge duals is given by 
\begin{align}
   \mathcal{L}_F 
   = -\frac{M_{\rm pl}^{D-2}}{4(p+2)!}
   \left( K^{IJ}(\Phi)\,F_I F_J 
   - \tilde K^{IJ}(\Phi)\,\star F_I F_J \right) \,.
\end{align}
Note that we lower and raise indices $I,J$ labeling different gauge fields using $K_{IJ}$. We define the electric and magnetic charges as
\begin{align}
    q_I &= \int \star F_I + \tilde K_{IJ} F^J  \,, \\
    p^J &= \int F^J \,,
\end{align}
where the integrals are over spheres that link with the brane. Note that the definition of electric charge is modified from the standard definition so that it remains quantized. The magnetic contribution to the electric charge is due to the Witten effect~\cite{Witten:1979ey}. In terms of these charges, the effective charge is defined as
\begin{align}
    q_{\rm eff}^2 
    = \frac{M_{\rm pl}^{D-2}}{2} \Big(
      q_I q_J\, K^{IJ}(\Phi) 
      + p^I p^J\, K_{IJ}
      \Big)\,.
\end{align}

Note that if the effective charge exhibits a runaway behavior—that is, if it does not have a minimum at a point in the moduli space, the resulting solution fails to produce a regular horizon. Instead, the scalar fields are driven to infinity, typically resulting in a singular horizon. Similarly, if the effective charge vanishes at a point in the moduli space, the resulting black hole will have zero area and the horizon would be singular. Such black holes are known as small black holes \cite{Sen:1994eb,Sen:1995in}.

A simple way to ensure a regular horizon is to consider a configuration where the gauge coupling does not depend on moduli. A well-known example is the black 3-brane in type IIB supergravity, for which the dilaton is constant and the gauge coupling does not run. More generally, black branes charged under moduli-dependent gauge fields can still have regular horizons, provided the function $q_{\rm eff}^2(\Phi)$ has a non-zero minimum in the moduli space. The D1/D5 system is a famous example of this situation, which is one of the reasons it was first used for the microscopic derivation of black hole entropy~\cite{Strominger:1996sh}. 

\subsection{Other Ricci-flat asymptotics}\label{exotic}

In subsection \ref{MinB}, we studied extremal black branes with Minkowski asymptotics, for which the transverse dimensions are given by a radial direction and a sphere. However, the transverse sphere can be replaced by more general manifolds. 

Consider extremal black brane geometries (where by ``extremal'' we mean that they share the same metric ansatz as in section \ref{MinB}) where the transverse sphere is replaced by a Sasaki--Einstein manifold $X$. To be more precise, consider extremal black 3-brane in type IIB supergravity, except that the transverse $S^5$ is replaced with $X^5$, a Sasaki--Einstein space with the same curvature. These are extremal black branes whose near-horizon geometry realizes the Klebanov--Witten vacua, which correspond to $\text{AdS}_5 \times X^5$ compactifications in type IIB string theory~\cite{Klebanov:1998hh}. Since Einstein equations are local, replacing the $S^5$ with any Einstein manifold would solve the local equations of motion. 
\begin{align}
ds_E^2 \;=\; \left(1 + \frac{L^4}{r^4}\right)^{-1/2}\,(-dt^2 + dx_1^2 + dx_2^2 + dx_3^2) 
\;+\; \left(1 + \frac{L^4}{r^4}\right)^{1/2}\,(dr^2 + r^2 ds_X^2),
\end{align}
where $L$ is a positive constant that sets the horizon scale and $ds_X$ is the metric on the Einstein Sasaki manifold with Ricci scalar curvature $\mathcal{R}_{X^5}=20$ so that it would match the Ricci scalar of the five sphere. 

To get the metric in the form of \eqref{metricform1}, we simply need to define 
\begin{align}
    \rho(r)&=\int \left(1 + \frac{L^4}{r^4}\right)^{1/4}dr\nonumber\\
    R(\rho)&=r\left(1 + \frac{L^4}{r^4}\right)^{1/4}\nonumber\\
    f(\rho)&=\left(1 + \frac{L^4}{r^4}\right)^{-1/2}\,.
\end{align}
and the metric becomes 
\begin{align}\label{metricform2}
    ds^2 \;=\; f(\rho)^2 \big[-dt^2 + d\vec{y}_p^2\big] \;+\; d\rho^2 \;+\; R(\rho)^2\, ds_{X}^2 \,.
\end{align}

This extremal black 3-brane is charged under the self-dual five-form flux $F_5$, just like the standard black 3-brane in ten-dimensional type IIB supergravity with Minkowski asymptotics. As explained in~\cite{Apers:2025pon}, this is equivalent to placing a black brane in a spacetime with the topology of $\mathbb{R}^4\times \mathbb{R}^{\geq 0}\times X$ where $\mathbb{R}^+\times X$ is a cone over $X$. In the case where $X$ is $S^5$, this spacetime is simply the Minkowski background. The state without any branes can be obtained by removing the fluxes from the solution, yielding a spacetime with a conical singularity.  

\subsection{AdS asymptotics}\label{AdSdS}

In this section, we study a further generalization of regular extremal black branes, where the spacetime asymptotics are no longer Ricci-flat. In particular, we consider black brane solutions whose asymptotics locally approach an AdS geometry.

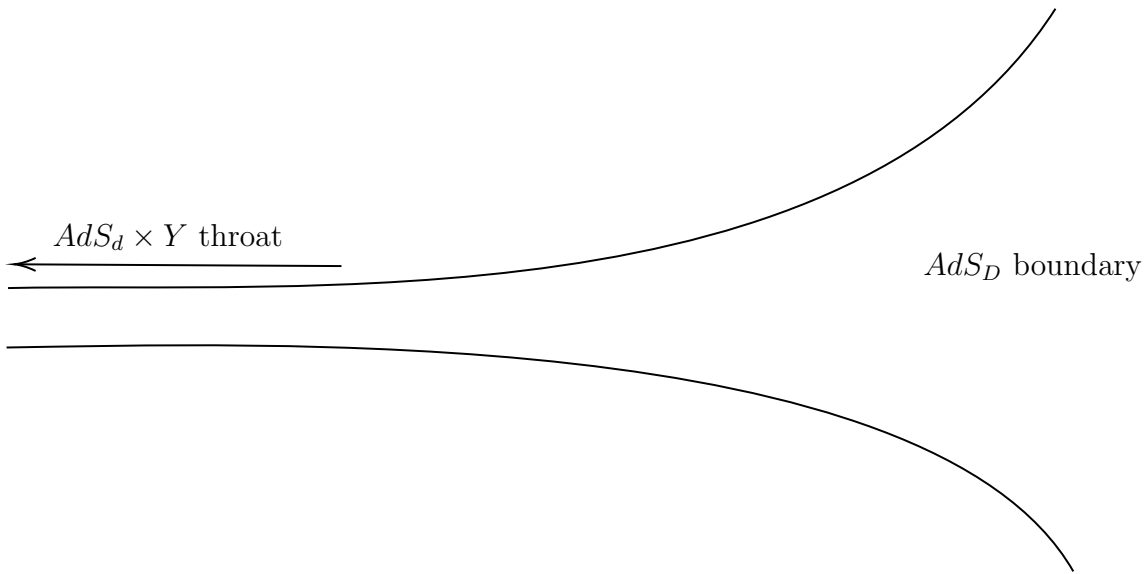
\begin{figure}[H]
    \centering

\tikzset{every picture/.style={line width=0.75pt}} 

\begin{tikzpicture}[x=0.75pt,y=0.75pt,yscale=-1,xscale=1]

\draw    (47.24,175) .. controls (188.73,171.94) and (472.5,195) .. (575.5,34) ;
\draw    (46.49,205) .. controls (164.87,203.21) and (511.5,191) .. (584.5,318) ;
\draw    (215.4,163.9) -- (52.23,163.01) ;
\draw [shift={(50.23,163)}, rotate = 0.31] [color={rgb, 255:red, 0; green, 0; blue, 0 }  ][line width=0.75]    (10.93,-3.29) .. controls (6.95,-1.4) and (3.31,-0.3) .. (0,0) .. controls (3.31,0.3) and (6.95,1.4) .. (10.93,3.29)   ;

\draw (68.4,141.01) node [anchor=north west][inner sep=0.75pt]   [align=left] {$\displaystyle AdS_{d} \times Y\ $throat};
\draw (507.4,156.01) node [anchor=north west][inner sep=0.75pt]   [align=left] {$\displaystyle AdS_{D} \ $boundary};

\end{tikzpicture}
    \caption{Extremal black brane in an asymptotically AdS$_D$ spacetime.}
    \label{AdSinAdS}
\end{figure}

The discussion in Section~\ref{MinB} about the near-horizon geometry of extremal branes applies more generally to extremal regular branes in any spacetime background. Therefore, the near-horizon geometry will still be a product of an AdS space and a compact space. Consequently, the spacetimes we are interested in interpolate between two AdS spaces of different dimensions (Figure~\ref{AdSinAdS}).
Let us review a concrete example of a geometry that interpolates between \( \text{AdS}_5 \times S^5 \) and \( \text{AdS}_3 \times Y \) in type IIB supergravity, where \( Y \) is a nontrivial fibration of \( S^5 \) over a Riemann surface \( \Sigma \)~\cite{Maldacena:2000mw}.
 Near the black brane (corresponding to $r \to \infty$), the metric takes the form:
\begin{align}
ds^2 \simeq 2^{-\frac{3}{2}} \sqrt{\Delta} \, ds_3^2 
& + 2^{-\frac{3}{2}} \sqrt{\Delta} \Bigg[ 
 \frac{\Delta}{y^2} (dx^2 + dy^2) 
+ 2\Delta\, d\theta^2 + 2\sin^2\theta\, d\phi_3^2\nonumber \\
& + 4\cos^2\theta \left( d\psi^2 + \sin^2\psi \left( d\phi_1 + \frac{1}{2y} dx \right)^2 + \cos^2\psi \left( d\phi_2 + \frac{1}{2y} dx \right)^2 \right) \Bigg],
\end{align}
where,
\begin{align}
ds_3^2 &= \frac{dr^2 - dt^2 + dz^2}{r^2}, ~~&\Delta = 1 + \sin^2\theta\,.
\end{align}
The angular coordinates $\theta, \psi, \phi_i$ (for $i = 1, 2, 3$) parametrize a squashed $S^5$, while $x$ and $y$ are coordinates on a Riemann surface. Note that $ds_3^2$ denotes the $AdS_3$ metric. It is useful to truncate the ten-dimensional equations of motion on the internal $S^5$ to obtain an effective five-dimensional theory. In this reduced theory, the geometric moduli describing the deformation of $S^5$ appear as scalar fields in five dimensions. More concretely, the non-trivial fibration of $S^5$ is captured by two scalar fields $\phi_1$ and $\phi_2$ (depending only on the radial coordinate $r$) and three Abelian gauge fields descending from the isometries of $S^5$ that act as rotations in the three orthogonal planes transverse to the brane. For more details on this perspective, see the Appendix of~\cite{Maldacena:2000mw}. 

The resulting five-dimensional metric takes the form~\cite{Benini:2013cda}:
\begin{align}
ds^2_5 = f(r)^2(-dt^2 + dz^2 + dr^2) + e^{2g(r) + 2h(x,y)} (dx^2 + dy^2),
\end{align}
with fluxes that thread the Riemann surface as
\begin{align}
F^I = -a^I\, e^{2h(x,y)}\, dx \wedge dy, \qquad I = 1, 2, 3,
\end{align}
where the constants $a^I$ determine the topological twist, and $h(x, y)$ determines the metric on the Riemann surface $\Sigma$. For instance, for a Riemann surface of genus $g > 1$, one typically chooses $h(x, y) = -\log y$.

Close to the brane ($r \rightarrow \infty$), the metric behaves as:
\begin{align}
f(r) \simeq \frac{L_{AdS_3}}{r}, \qquad g(r) \to \text{const},
\end{align}
which describes an $AdS_3 \times \Sigma$ geometry. Far from the brane ($r \rightarrow 0$), the solution asymptotes to a deformed $AdS_5$:
\begin{align}
ds^2 \simeq L_{AdS_5}^2 \frac{-dt^2 + dz^2 + e^{2h(x,y)}(dx^2 + dy^2) + dr^2}{r^2}.
\end{align}

By defining the proper radial distance
\begin{align}
\rho = -\int_\infty^r f(r') dr',
\end{align}
we can rewrite the metric as
\begin{align}\label{5DRS}
ds_5^2 = f(\rho)^2(-dt^2 + dz^2) + d\rho^2 + e^{2g(\rho) + 2h(x,y)}(dx^2 + dy^2),
\end{align}
where the horizon of the black brane (with $AdS_3$ near-horizon geometry) is located at $\rho \rightarrow -\infty$, and the asymptotic region (where the geometry becomes locally $AdS_5$) lies at $\rho \rightarrow \infty$.

The asymptotic behavior of the warp factors is:
\begin{align}
&\rho \rightarrow -\infty: \quad f(\rho) \sim e^{\rho / L_{AdS_3}}, \qquad g(\rho) \rightarrow \text{const}, \\
&\rho \rightarrow \infty: \quad f(\rho) \sim e^{\rho / L_{AdS_5}}, \qquad g(\rho) \sim \frac{\rho}{L_{AdS_5}}.
\end{align}

These supergravity solutions represent RG flows in their holographic duals. The boundary of the spacetime is asymptotically AdS, deformed by a term that vanishes at the boundary. Thus, UV correlators exhibit conformal symmetry, as in the undeformed $AdS_5$ geometry. However, as we move inward along the radial direction (interpreted as flowing to the IR in the dual theory), the geometry transitions to an $AdS_3$ space, implying the emergence of a 2D conformal symmetry.

This corresponds to a quantum field theory that flows from a four-dimensional conformal theory in the UV to a two-dimensional conformal theory in the IR. In the example mentioned above, the holographic dual is $\mathcal{N}=4$ SYM compactified on a Riemann surface. The six scalars in SYM can be organized into three complex scalar fields, which may have nontrivial bundles over the Riemann surface. In certain cases, these preserve supersymmetry. The introduction of a Riemann surface breaks conformal invariance due to its intrinsic scale, but supersymmetry can remain intact. Consequently, the theory behaves as a 4D $\mathcal{N}=4$ SCFT in the UV, and flows to a 2D SCFT in the IR.

\section{Existence of the Decoupling Limit}\label{dec}

In this section, we review how low-energy modes in the near-horizon geometry of a black brane, as well as excitations of branes supporting a worldvolume theory, decouple from the gravitational dynamics in the ambient spacetime. This decoupling is a crucial ingredient for the emergence of holography~\cite{Maldacena:1997re}, where the gravitational theory in the throat can be equivalently described by the worldvolume theory on a brane.

\subsection{Decoupling of Near-Horizon Geometries}\label{decopblack}

In the background of a static black brane, the energies of quasi-stationary modes vary with position due to gravitational redshift. This redshift is governed by the $tt$ component of the metric: as a mode moves away from the black brane, its locally measured energy decreases.

Let us begin by focusing on extremal black branes with regular horizons in asymptotically Minkowski spacetime, as reviewed in Section~\ref{MinB}. 

\subsubsection*{(i) Extremal branes in Minkowski with regular horizons}

The metric for such branes is
\begin{align}
    ds^2 \;=\; f(\rho)^2 \big[-dt^2 + d\vec{y}_p^2\big] + d\rho^2 + R(\rho)^2\, ds^2_{D-p-2},
\end{align}
where $f(\rho) \to \text{const.}$ as $\rho \to \infty$, and $f(\rho) \to 0$ near the horizon at $\rho_H = -\infty$. We will generalize to other asymptotics later.

The energy of a given mode redshifts as $f(\rho)^{-1}$, and hence the energy measured at spatial infinity decreases as the mode travels away from the brane. The AdS throat lies at $\rho \to -\infty$, where $f(\rho) \sim e^{\rho / L_{\text{AdS}}}$. Suppose a mode has local energy $E$ deep in the throat at $\rho_i \ll -L$. Then its redshifted energy as measured at infinity is
\begin{align}
    E_{\infty} = \lim_{\rho_o \to \infty} \frac{f(\rho_i)}{f(\rho_o)} E.
\end{align}

Thus, modes localized deeper in the throat ($\rho_i \to -\infty$) appear increasingly soft to an asymptotic observer. Equivalently, one can hold $\rho_i$ fixed and consider modes of smaller energy $E$. The specification of a mode by the pair $(\rho_i, E)$ contains redundant information, since every mode can be characterized by its energy evaluated at any location. It is the invariant product $f(\rho_i) E$ that uniquely determines the physical mode. This is analogous to the concept of \textit{comoving energy} in cosmology, which provides a time-invariant notion of energy for weakly coupled perturbations. Just as the scale factor in cosmology accounts for the redshift due to cosmological expansion, here the redshift factor $f(\rho_i)$ plays a similar role.

\begin{figure}[H]
    \centering

  
\tikzset {_kbo8ijs06/.code = {\pgfsetadditionalshadetransform{ \pgftransformshift{\pgfpoint{0 bp } { 0 bp }  }  \pgftransformrotate{0 }  \pgftransformscale{2 }  }}}
\pgfdeclarehorizontalshading{_2iidegznf}{150bp}{rgb(0bp)=(1,1,0);
rgb(37.5bp)=(1,1,0);
rgb(62.5bp)=(0,0.5,0.5);
rgb(100bp)=(0,0.5,0.5)}
\tikzset{every picture/.style={line width=0.75pt}} 

\begin{tikzpicture}[x=0.75pt,y=0.75pt,yscale=-1,xscale=1]

\draw    (26.24,141) .. controls (167.73,137.94) and (451.5,161) .. (554.5,0) ;
\draw    (25.49,171) .. controls (143.87,169.21) and (490.5,157) .. (563.5,284) ;
\draw  [color={rgb, 255:red, 74; green, 144; blue, 226 }  ,draw opacity=1 ][line width=1.5]  (32,157.5) .. controls (32.82,161.34) and (33.6,165) .. (34.5,165) .. controls (35.4,165) and (36.18,161.34) .. (37,157.5) .. controls (37.82,153.66) and (38.6,150) .. (39.5,150) .. controls (40.4,150) and (41.18,153.66) .. (42,157.5) .. controls (42.82,161.34) and (43.6,165) .. (44.5,165) .. controls (45.4,165) and (46.18,161.34) .. (47,157.5) .. controls (47.82,153.66) and (48.6,150) .. (49.5,150) .. controls (50.4,150) and (51.18,153.66) .. (52,157.5) .. controls (52.82,161.34) and (53.6,165) .. (54.5,165) .. controls (55.4,165) and (56.18,161.34) .. (57,157.5) .. controls (57.82,153.66) and (58.6,150) .. (59.5,150) .. controls (60.4,150) and (61.18,153.66) .. (62,157.5) .. controls (62.82,161.34) and (63.6,165) .. (64.5,165) .. controls (65.4,165) and (66.18,161.34) .. (67,157.5) .. controls (67.82,153.66) and (68.6,150) .. (69.5,150) .. controls (70.4,150) and (71.18,153.66) .. (72,157.5) .. controls (72.82,161.34) and (73.6,165) .. (74.5,165) .. controls (75.4,165) and (76.18,161.34) .. (77,157.5) .. controls (77.82,153.66) and (78.6,150) .. (79.5,150) .. controls (80.4,150) and (81.18,153.66) .. (82,157.5) .. controls (82.82,161.34) and (83.6,165) .. (84.5,165) .. controls (85.4,165) and (86.18,161.34) .. (87,157.5) .. controls (87.82,153.66) and (88.6,150) .. (89.5,150) .. controls (90.4,150) and (91.18,153.66) .. (92,157.5) .. controls (92.82,161.34) and (93.6,165) .. (94.5,165) .. controls (95.4,165) and (96.18,161.34) .. (97,157.5) .. controls (97.82,153.66) and (98.6,150) .. (99.5,150) ;
\draw  [color={rgb, 255:red, 208; green, 2; blue, 27 }  ,draw opacity=1 ][line width=1.5]  (465.5,147.5) .. controls (468.76,150.83) and (471.88,154) .. (475.5,154) .. controls (479.12,154) and (482.24,150.83) .. (485.5,147.5) .. controls (488.76,144.17) and (491.88,141) .. (495.5,141) .. controls (499.12,141) and (502.24,144.17) .. (505.5,147.5) .. controls (508.76,150.83) and (511.88,154) .. (515.5,154) .. controls (519.12,154) and (522.24,150.83) .. (525.5,147.5) .. controls (528.76,144.17) and (531.88,141) .. (535.5,141) .. controls (539.12,141) and (542.24,144.17) .. (545.5,147.5) .. controls (548.76,150.83) and (551.88,154) .. (555.5,154) .. controls (559.12,154) and (562.24,150.83) .. (565.5,147.5) .. controls (568.76,144.17) and (571.88,141) .. (575.5,141) .. controls (579.12,141) and (582.24,144.17) .. (585.5,147.5) .. controls (588.76,150.83) and (591.88,154) .. (595.5,154) .. controls (597.63,154) and (599.59,152.9) .. (601.5,151.32) ;
\draw [shading=_2iidegznf,_kbo8ijs06]   (68,248) -- (466.5,249) ;
\draw [shift={(468.5,249)}, rotate = 180.14] [color={rgb, 255:red, 0; green, 0; blue, 0 }  ][line width=0.75]    (10.93,-3.29) .. controls (6.95,-1.4) and (3.31,-0.3) .. (0,0) .. controls (3.31,0.3) and (6.95,1.4) .. (10.93,3.29)   ;

\draw (22.4,111.01) node [anchor=north west][inner sep=0.75pt]   [align=left] {AdS throat};
\draw (504.4,101.01) node [anchor=north west][inner sep=0.75pt]   [align=left] {Asymptotic boundary};
\draw (170,259) node [anchor=north west][inner sep=0.75pt]   [align=left] {Redshifting of gravitational modes};
\draw (26,184) node [anchor=north west][inner sep=0.75pt]   [align=left] {Position:\\Energy: \ };
\draw (91,184.4) node [anchor=north west][inner sep=0.75pt]    {$ \begin{array}{l}
\rho =\rho _{i}\\
E
\end{array}$};
\draw (567,169.4) node [anchor=north west][inner sep=0.75pt]    {$ \begin{array}{l}
\rho =\rho _{o}\\
\end{array}$};
\draw (502,169) node [anchor=north west][inner sep=0.75pt]   [align=left] {Position:\\\\Energy: \ };
\draw (563,198.4) node [anchor=north west][inner sep=0.75pt]    {$E\frac{f( \rho _{i})}{f( \rho _{o})}$};

\end{tikzpicture}
\caption{Modes with local energy $E$ in the throat at $\rho = \rho_i$ experience a gravitational redshift as they propagate to the asymptotic region $\rho_o \gg L$.}
    \label{Redshift}
\end{figure}
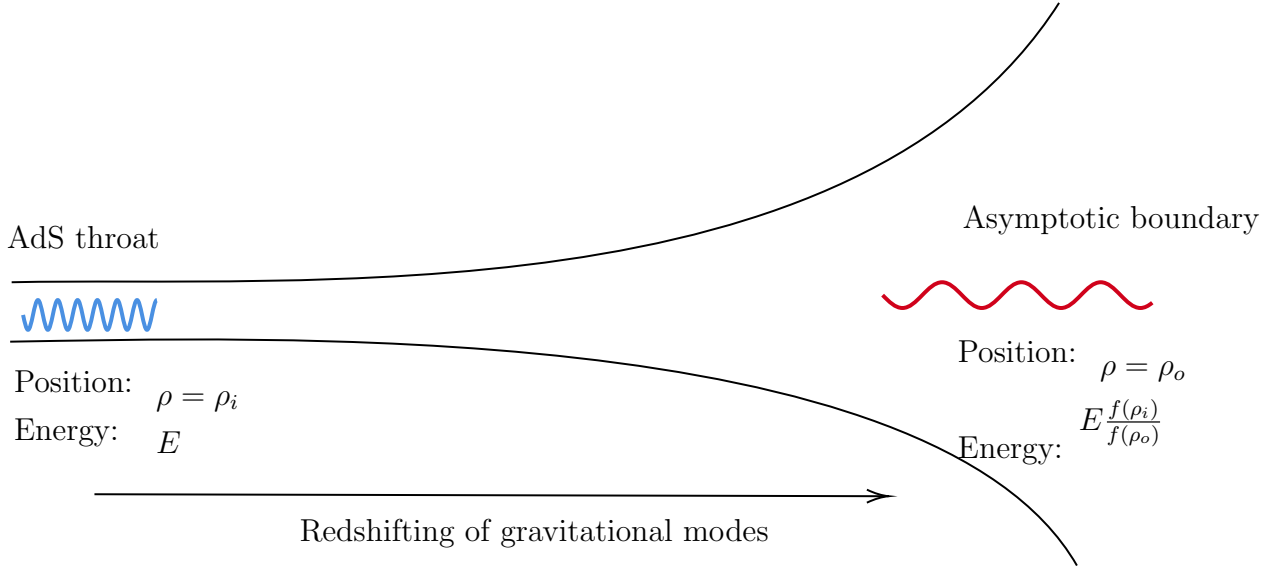
In gravitational theories, the energy of a mode also governs its coupling to gravity. For instance, perturbations become strongly coupled at infinity when $E_\infty \gg M_{\rm pl}$, since the Einstein term produces radiative higher-derivative corrections that are suppressed by $M_{\rm pl}$. More precisely, this coupling is set by
\[
g_{\rm{grav.}}=\frac{E_\infty}{\Lambda_s},
\]
where $\Lambda_s$ is the relevant cutoff scale for gravitational interactions (which could be set by the string scale or the higher-dimensional Planck scale, as we reviewed in Section~\ref{AdSdist}). Therefore, the condition for weak gravitational coupling at infinity becomes
\begin{align}
    \text{\rm Weak coupling condition:} \quad \frac{E_{\infty}}{\Lambda_s} \ll 1.
\end{align}

Substituting the redshifted energy, we find the effective coupling of a mode with local energy $E$ at position $\rho_i$ is
\begin{align}
    g_{\text{grav.}} = \lim_{\rho_o \to \infty} \frac{f(\rho_i)}{f(\rho_o)} \cdot \frac{E}{\Lambda_s}.
\end{align}
Thus, by considering low-energy modes (or modes deeper in the throat), the coupling to asymptotic gravity becomes arbitrarily small, and the mode effectively decouples from the external spacetime.

\begin{align}
    \text{\rm Decoupled modes:} \quad g_{\text{grav.}} \ll 1.
\end{align}
Now let us consider other types of branes discussed in Section~\ref{Ext}. These include:  
(i) branes with singular horizons,  
(ii) branes with non-Minkowski Ricci-flat asymptotics, and  
(iii) branes with AdS asymptotics.  

In each case, we demonstrate the key features with a representative example, some of which were reviewed in Section~\ref{Ext}. However, the arguments apply more generally.

\subsubsection*{(ii) Singular branes}

As discussed in Section~\ref{sing}, extremal branes charged under gauge fields with moduli-dependent couplings that vanish at infinite distance in moduli space exhibit singular horizons. Since the horizon is singular, the near-horizon geometry is no longer AdS. Nevertheless, such branes can still have holographic descriptions. A well-known example is \emph{little string theory}\cite{Seiberg:1997zk}, where the near-horizon geometry of a stack of NS5-branes (described by a linear dilaton background \cite{Callan:1991at,Seiberg:1997zk}) decouples from the gravity away from the brane and is dual to a six-dimensional non-local theory known as little string theory~\cite{Aharony:1998ub}.

The Einstein-frame metric of NS5-branes in type II string theory is given by:
\begin{align}
ds^2_{\text{Einstein}} = H(r)^{-1/4} \left( -dt^2 + dx_1^2 + \cdots + dx_5^2 \right) + H(r)^{3/4} \left( dr^2 + r^2 d\Omega_3^2 \right),
\end{align}
where the harmonic function takes the form
\[
H(r) = 1 + \frac{L^2}{r^2}\,.
\]

The canonically normalized dilaton varies with position as
\begin{align}
e^{\sqrt{8}\phi} \propto H(r)\,,
\end{align}
which diverges near the horizon $r \rightarrow 0$ and asymptotes to a constant as $r \rightarrow \infty$. 

To write the metric in the form of Eq.~\eqref{metricform1}, define the proper radial coordinate:
\begin{align}
\rho = \int dr\, H(r)^{3/8}.
\end{align}
In terms of $\rho$, the metric becomes:
\begin{align}\label{NS5}
ds^2_{\text{Einstein}} = f(\rho)^2\left( -dt^2 + dx_1^2 + \cdots + dx_5^2 \right) + d\rho^2 + R(\rho)^2 d\Omega_3^2,
\end{align}
where
\begin{align}
& \rho \rightarrow 0: \quad f(\rho) \sim \left( \frac{\rho}{L} \right)^2, \quad R(\rho) \sim L\left( \frac{\rho}{L} \right)^6, \\
& \rho \rightarrow \infty: \quad f(\rho) \rightarrow 1, \quad R(\rho) \simeq \rho\,.
\end{align}

As $\rho \rightarrow 0$, the curvature diverges, and the effective field theory description breaks down. Thus, the near-horizon region is not an AdS space but is instead described by a linear dilaton background.

Despite the singularity, one can still define the gravitational coupling of a mode located at $\rho_i$ by:
\begin{align}
g_{\text{grav}} = \lim_{\rho_o \to \infty} \frac{f(\rho_i)}{f(\rho_o)} \cdot \frac{E}{M_s(\rho_o)}\,,
\end{align}
where $M_s(\rho_o)$ is the local string scale at position $\rho_o$, which depends on the dilaton in Einstein frame.

As long as $g_{\text{grav}} \ll 1$, the corresponding mode in the throat is effectively decoupled from ten-dimensional gravity. Hence, just like in the case of regular extremal black branes, the decoupling of low-energy modes is given by the condition $g_{\rm{grav.}}\ll1$.

\subsubsection*{(iii) Non-Minkowski Ricci-flat asymptotics}

Now let us consider black branes in spacetimes that are asymptotically Ricci-flat and locally resemble Minkowski space, but globally are not Minkowski. An example of such a geometry was discussed in Section~\ref{exotic}, corresponding to a black brane whose near-horizon geometry realizes the Klebanov--Witten AdS vacua~\cite{Klebanov:1998hh}. These vacua arise as Freund--Rubin compactifications~\cite{Freund:1980xh} of type IIB supergravity on $AdS_5 \times X$, where $X$ is a five-dimensional Einstein manifold. The metric takes the form:
\begin{align}\label{metricform3}
    ds^2 = f(\rho)^2 \left[-dt^2 + d\vec{y}_p^2\right] + d\rho^2 + R(\rho)^2\, ds_X^2,
\end{align}
where $ds_X^2$ is the metric on the Einstein manifold $X^5$, and the functions $f(\rho)$ and $R(\rho)$ are defined through the following change of variables:
\begin{align}
    \rho(r) &= \int_0^r \left(1 + \frac{L^4}{r'^4}\right)^{1/4} dr', \\
    R(\rho) &= r \left(1 + \frac{L^4}{r^4} \right)^{1/4}, \\
    f(\rho) &= \left(1 + \frac{L^4}{r^4} \right)^{-1/2}.
\end{align}

The function $f(\rho)$ in these backgrounds is identical to that of the black 3-brane in type IIB supergravity with Minkowski asymptotics which corresponds to $X=S^5$. Since the redshift of modes outside the AdS throat (as $\rho \to -\infty$) is controlled by $f(\rho)$, the decoupling analysis proceeds identically to the Minkowski case.

In other words, weak gravitational coupling of a mode depends only on the local value of the metric component $g_{tt}$, which is determined by solving the Einstein equations locally and is thus insensitive to the global structure of the spacetime. As in the previous examples, the coupling of a mode to 10d gravity at infinity is given by:
\begin{align}
    g_{\text{grav}} = \lim_{\rho_o \to \infty} \frac{f(\rho_i)}{f(\rho_o)} \cdot \frac{E}{\Lambda_s},
\end{align}
and the decoupling condition becomes:
\begin{align}
    g_{\text{grav}} \ll 1.
\end{align}

Therefore, even though the spacetime is not globally Minkowski, as long as the redshift profile away from the brane is the same, the modes near the brane can still decouple from the gravity far away.

\subsubsection*{(iv) Branes with AdS asymptotics}

Here we consider extremal black branes in AdS spacetimes, which, as we explained in Section~\ref{AdSdS}, can be viewed as geometries interpolating between AdS spaces of different dimensions. For concreteness, we revisit the example discussed in Section~\ref{AdSdS} which interpolates between $AdS_3$ near the horizon and $AdS_5$ in the asymptotic region.

The five-dimensional metric for this geometry is given by:
\begin{align}
ds_5^2 = f(\rho)^2(-dt^2 + dz^2) + d\rho^2 + e^{2g(\rho) + 2h(x,y)}(dx^2 + dy^2),
\end{align}
with asymptotic behavior:
\begin{align}
\rho \rightarrow -\infty: &\quad f(\rho) \sim e^{\rho / L_{AdS_3}}, \qquad g(\rho) \rightarrow \text{const}, \\
\rho \rightarrow \infty: &\quad f(\rho) \sim e^{\rho / L_{AdS_5}}, \qquad g(\rho) \sim \frac{\rho}{L_{AdS_5}}.
\end{align}

A mode with local energy \(E\) located deep in the $AdS_3$ throat at \(\rho = \rho_i \ll -L_{AdS_3}\) appears to an observer far from the brane at \(\rho = \rho_o \gg L_{AdS_5}\) to have a redshifted energy given by $E \frac{f(\rho_i)}{f(\rho_o)}$. Now if we evaluate this energy in the asymtptotic of the spacetime, we find
\begin{align}
\frac{E_{\infty}}{M_{\rm pl,5}} = \lim_{\rho_o \rightarrow \infty} \frac{E}{M_{\rm pl,5}} \frac{f(\rho_i)}{f(\rho_o)} \sim \frac{E}{M_{\rm pl,5}}\, \exp\left(-\frac{\rho_o}{L_{AdS_5}}\right) \rightarrow 0.
\end{align}

Thus, such modes decouple from gravity in the asymptotic $AdS_5$ region. To ensure that the energy is sufficiently small to also decouple from gravity in the intermediate and bulk regions of $AdS_5$, one can either reduce \(E\) or localize the mode deeper in the $AdS_3$ throat. Both approaches lead to decoupling from the full bulk gravitational dynamics.

\subsection{Decoupling of Branes}\label{decopbrane}

In the previous section, we considered extremal black branes as fully gravitational solutions. However, in many scenarios within string theory, branes cannot be described as black branes. This is because, before reaching the would-be horizon region, the curvature can exceed the quantum gravity cutoff (e.g., the species scale), rendering the low-energy gravitational description invalid. In such cases, the region near the brane must be replaced by a non-gravitational defect theory that admits a microscopic description localized on the brane’s worldvolume.

This is the higher-dimensional analog of why we do not describe light particles as black holes. Although any neutral particle does gravitationally backreact on the spacetime and asymptotically produces a Schwarzschild-like geometry, the radius of such a black hole is so small that classical general relativity is no longer reliable in its vicinity. In particular, quantum gravitational effects become important well before reaching the would-be horizon. Hence, we do not interpret such particles as black holes.

When we refer to the \textit{worldvolume theory} of a brane, we are implicitly identifying the degrees of freedom that decouple from bulk gravity in the low-energy limit and become localized on the brane. A familiar example is the system of \( N \gg 1 \) D3-branes. In the regime \( N^{-1/2} \ll g_s \ll 1 \), the gravitational description near the horizon is valid, and the black brane description applies. However, in the opposite regime \( g_s \ll N^{-1/2} \), the appropriate picture is that of the D3-branes whose worldvolume theory is given by \( \mathcal{N} = 4 \) SYM at low energies. 

Crucially, the theory should not be viewed as a direct sum of 10-dimensional supergravity and 4-dimensional \( \mathcal{N} = 4 \) SYM. The two sectors are only decoupled at sufficiently low energies. Specifically, decoupling occurs in the infrared limit, where massless degrees of freedom on the brane interact only weakly with the 10d gravitational bulk. In this context, \textit{low energy} means energies well below the string scale \( M_s \), which controls the strength of coupling to 10d gravity.

To see this concretely, consider a generic neutral perturbation on the brane. Eventually, this perturbation decays into bulk modes that propagate away from the brane. That is, all modes localized on the brane necessarily have tails in spacetime that extend into the asymptotic region far away from the brane. The strength of the coupling between the brane degrees of freedom and the bulk modes is governed by the ratio \( E / M_s \), where \( E \) is the energy of the emitted mode as measured far from the brane, and \( M_s \) is the local string scale. If this ratio is large, the gravitational backreaction of the perturbation becomes significant even in the asymptotic region, making it inappropriate to regard the excitation as a brane-localized degree of freedom. In such cases, the perturbation is better understood as a bulk gravitational excitation.
 Conversely, for
\begin{align}
    \frac{E_\infty}{\Lambda_s} \ll 1\,,
\end{align}
the coupling is weak, and the brane degrees of freedom effectively decouple from the bulk gravity. Here \( \Lambda_s \) is the species scale that controls higher-derivative corrections, typically set by the string scale. Thus, ensuring that \( E_\infty \ll \Lambda_s \) is essential for the emergence of a decoupled, non-gravitational worldvolume theory on the brane. Importantly, this argument applies universally to all branes, regardless of the asymptotic structure of the ambient spacetime. The reason is that decoupling is fundamentally controlled by the behavior of energy scales far away from the brane, where the spacetime is well-approximated by the black brane geometry. This approximation enables us to apply the analysis of the previous subsection, which demonstrated that the decoupling argument holds irrespective of the global asymptotics of the spacetime.

There is a broader statement here: to realize a superconformal field theory (SCFT) as the infrared IR limit of a brane in string theory, it is crucial that gravity becomes weakly coupled at large distances. This allows one to meaningfully take the IR limit by probing large-distance physics without encountering strong gravitational interactions. In this sense, the existence of a decoupling limit is not merely a property of the near-horizon geometry, but a feature that hinges on the behavior of gravity in the asymptotic region.

\subsection{Existence condition}

In subsections~\ref{decopblack} and~\ref{decopbrane}, we showed that, for the near-horizon geometry of a black brane to decouple from the asymptotic gravity or for the worldvolume theory of a brane in string theory to decouple from gravity, we must consider energies $E_\infty$ (as measured from infinity) that satisfy
\begin{align}
    \frac{E_{\infty}}{\Lambda_s(\infty)} 
    = \lim_{\rho_o \rightarrow \infty} \frac{E}{\Lambda_s(\rho_o)}\, \frac{f(\rho_i)}{f(\rho_o)}  \ll 1\,.
\end{align}

It is easy to see that the existence of a decoupling limit is highly sensitive to the behavior of the ratio $E/\Lambda_s$ far away from the brane. We distinguish the following possibilities:

\begin{itemize}
    \item \textbf{Case I:} 
    $\displaystyle \lim_{\rho \to \infty} \frac{M_{\rm pl}}{f(\rho) \Lambda_s(\rho)} < \infty$.  
    \\
    In this case, as we move to lower energies (corresponding to excitations deeper inside the throat), such modes decouple from higher-dimensional gravity far from the brane. The limit may either vanish or approach a finite constant. For example, as explained in subsection~\ref{decopblack}, for extremal black branes in AdS spaces the ratio vanishes, while for extremal black branes in Minkowski space, the ratio asymptotes to a constant.
    
    \item \textbf{Case II:} 
    $\displaystyle \lim_{\rho \to \infty} \frac{M_{\rm pl}}{f(\rho) \Lambda_s(\rho)} \to \infty$.  
    \\
    In this case, no modes can decouple from gravity—not even gravity itself. Gravitational perturbations become strongly coupled asymptotically, implying a breakdown of the semiclassical description.
\end{itemize}

For holography to be well-defined, we require Case~I. This condition is realized in standard examples, including the linear dilaton backgrounds dual to little string theory, as discussed in subsection~\ref{decopblack}. We therefore arrive at the following statement:

\begin{tcolorbox}[colback=gray!10, colframe=black, title=Decoupling condition]
\medskip

\noindent
\textit{In order for modes in the near-horizon geometry of a black brane, or the worldvolume theory on a brane, to decouple from gravity in the higher-dimensional ambient theory, the asymptotics of the spacetime must satisfy}
\begin{align}\label{deccond}
\lim_{\rho \to \infty} \frac{M_{\rm pl}}{f(\rho)\Lambda_s(\rho)} \;<\;\infty \,.
\end{align}

\end{tcolorbox}

We refer to this as the \emph{decoupling condition}.

\section{Reconstructing black brane from the AdS}\label{DC}

In this section, we consider a generic theory that has an AdS solution and construct the black brane solution that realizes that AdS in the near horizon limit. 

\subsection{From AdS to black brane}\label{uplift}

Consider a theory with an $\mathrm{AdS}_d$ solution. We assume that this theory includes some scalar fields, that may be massive. This assumption is correct for AdS solutions that arise from string theory, as they result from compactifications of a higher-dimensional theory. The geometric moduli of the compactification manifold become scalar fields in the lower-dimensional effective theory. 

These scalar fields are governed by an effective scalar potential, $V(\phi)$, which arises from fluxes, branes, and curvature in the compactification. In Freund--Rubin compactifications~\cite{Freund:1980xh}, for example, the higher-dimensional theory is compactified on a compact internal manifold threaded with fluxes. This leads to a lower-dimensional theory with a scalar potential that depends on the moduli fields controlling the shape and size of the internal space.

The $\mathrm{AdS}_d$ vacuum corresponds to the configuration in which all scalar fields are fixed at a critical point $\phi = \phi_0$ of the potential, where $V(\phi_0) < 0$. At this critical point, the scalar fields are constant across spacetime, and the geometry is exactly AdS.

We are interested in constructing black brane solutions in the lower-dimensional theory that interpolate between this AdS vacuum near the horizon and some other geometry far away. To this end, it suffices to focus on the truncated equations of motion involving only the scalar fields and the metric. We neglect other fields in the theory, assuming they vanish.

In particular, suppose the scalar potential $V(\phi)$ has a global minimum at $\phi_0$ with negative vacuum energy $\Lambda = V(\phi_0) < 0$, and that $V(\phi) \to 0$ as $\phi \to \infty$. This behavior allows for the construction of a black brane solution where the scalar field $\phi(r)$ runs from $\phi_0$ near the horizon (realizing the $\mathrm{AdS}_d$ geometry) to large values far away from the brane, where the potential effectively vanishes.

We will see that under this setup, the black brane solution takes a simple and universal form governed by the behavior of the scalar potential and the geometry induced by the running of $\phi(r)$.
 The corresponding black brane solution has metric
\begin{align}\label{DCm}
   ds^2 = a(x)^2 \left( -dt^2 + \sum_{i=1}^{d-2} d\sigma_i^2 \right) + dx^2\,, 
\end{align}
and a scalar field profile $\phi(x)$ satisfying
\begin{align}\label{DCs}
    \phi(x) \;\to\; \phi_0 \quad \text{as} \quad x \to -\infty, 
\qquad 
\phi(x) \;\to\; \infty \quad \text{as} \quad x \to +\infty\,.
\end{align}

For example, in the $\mathrm{AdS}_5 \times S^5$ compactification, the relevant scalar field is the canonically normalized volume modulus of the $S^5$. The scalar potential---receiving contributions from both fluxes and internal curvature---attains a minimum at some fixed radius of $S^5$. In the solution described above, the volume modulus of the $S^5$ approaches its stabilized value as $x \to -\infty$, while diverging as $x \to +\infty$. This solution arises from the dimensional reduction of a black 3-brane carrying $N$ units of charge under the RR 4-form gauge potential.

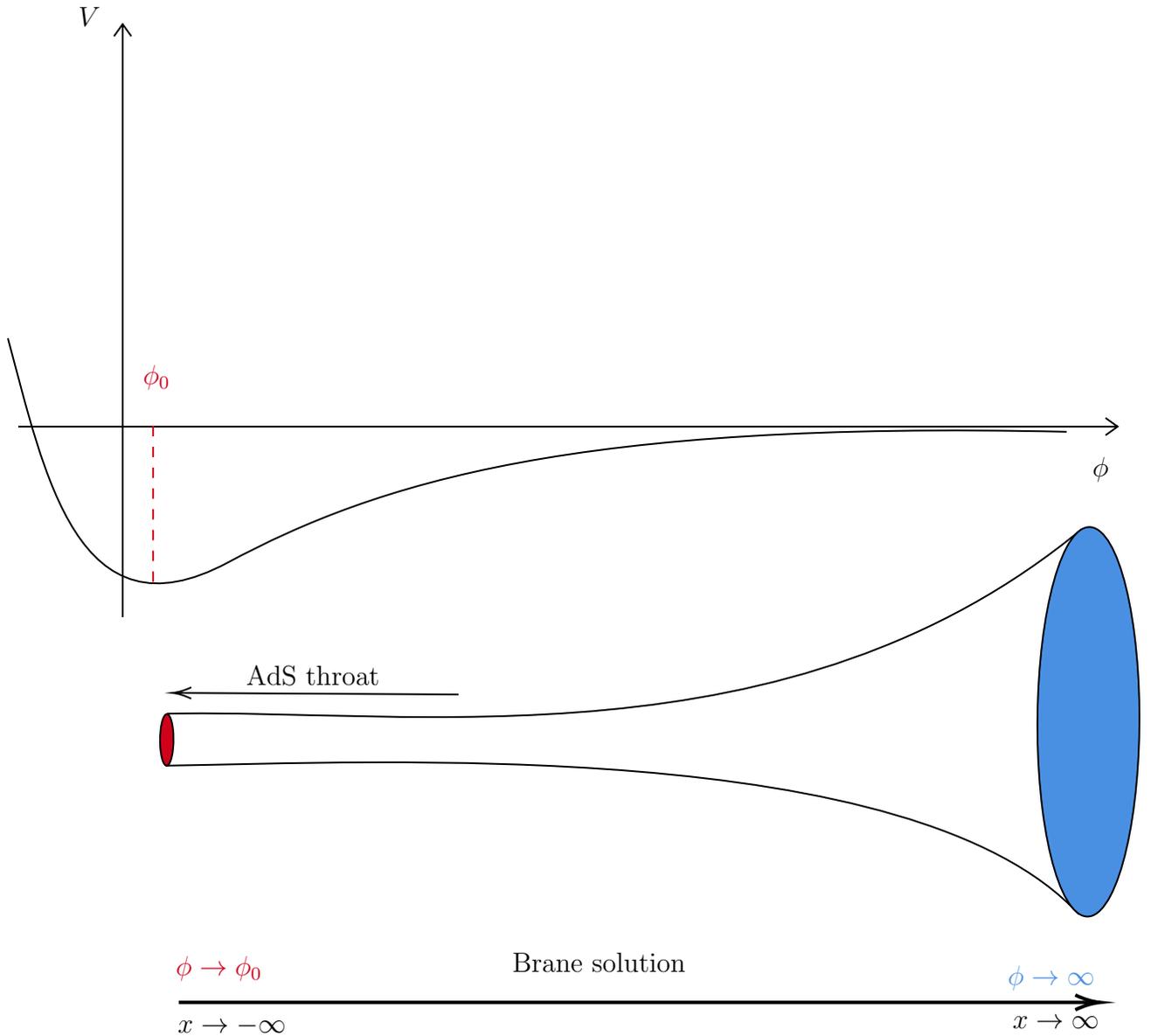
\begin{figure}[H]

  
\tikzset {_lleob0273/.code = {\pgfsetadditionalshadetransform{ \pgftransformshift{\pgfpoint{0 bp } { 0 bp }  }  \pgftransformrotate{0 }  \pgftransformscale{2 }  }}}
\pgfdeclarehorizontalshading{_vvto1l20j}{150bp}{rgb(0bp)=(1,1,0);
rgb(37.5bp)=(1,1,0);
rgb(62.5bp)=(0,0.5,0.5);
rgb(100bp)=(0,0.5,0.5)}
\tikzset{every picture/.style={line width=0.75pt}} 

\begin{tikzpicture}[x=0.75pt,y=0.75pt,yscale=-1,xscale=1]

\draw  (8.5,268) -- (642.91,268)(68.73,35.64) -- (68.73,378) (635.91,263) -- (642.91,268) -- (635.91,273) (63.73,42.64) -- (68.73,35.64) -- (73.73,42.64)  ;
\draw    (2.5,217) .. controls (23.5,295) and (42.5,393) .. (129.42,346.47) .. controls (216.34,299.95) and (329.5,264.47) .. (613.5,271) ;
\draw [color={rgb, 255:red, 208; green, 2; blue, 27 }  ,draw opacity=1 ] [dash pattern={on 4.5pt off 4.5pt}]  (86.29,357.62) -- (86.29,267.44) ;
\draw [shading=_vvto1l20j,_lleob0273][line width=1.5]    (101.13,600.44) -- (629.67,600.44) ;
\draw [shift={(632.67,600.44)}, rotate = 180] [color={rgb, 255:red, 0; green, 0; blue, 0 }  ][line width=1.5]    (14.21,-4.28) .. controls (9.04,-1.82) and (4.3,-0.39) .. (0,0) .. controls (4.3,0.39) and (9.04,1.82) .. (14.21,4.28)   ;
\draw    (94.39,433.74) .. controls (235.88,430.68) and (453.93,466.42) .. (622.38,327.03) ;
\draw    (93.64,463.74) .. controls (212.02,461.95) and (525.05,445.87) .. (620.51,549.51) ;
\draw  [fill={rgb, 255:red, 74; green, 144; blue, 226 }  ,fill opacity=1 ] (599.46,488.58) .. controls (592.49,433.01) and (598.74,365.51) .. (613.43,337.81) .. controls (628.12,310.11) and (645.68,332.71) .. (652.65,388.28) .. controls (659.62,443.85) and (653.37,511.35) .. (638.68,539.05) .. controls (623.99,566.74) and (606.43,544.15) .. (599.46,488.58) -- cycle ;
\draw  [fill={rgb, 255:red, 208; green, 2; blue, 27 }  ,fill opacity=1 ] (90.67,455.5) .. controls (89.75,448.13) and (90.57,439.19) .. (92.52,435.52) .. controls (94.47,431.86) and (96.79,434.85) .. (97.71,442.21) .. controls (98.64,449.57) and (97.81,458.51) .. (95.86,462.18) .. controls (93.92,465.85) and (91.59,462.86) .. (90.67,455.5) -- cycle ;
\draw    (262.55,422.64) -- (99.38,421.75) ;
\draw [shift={(97.38,421.74)}, rotate = 0.31] [color={rgb, 255:red, 0; green, 0; blue, 0 }  ][line width=0.75]    (10.93,-3.29) .. controls (6.95,-1.4) and (3.31,-0.3) .. (0,0) .. controls (3.31,0.3) and (6.95,1.4) .. (10.93,3.29)   ;

\draw (41.7,24.17) node [anchor=north west][inner sep=0.75pt]    {$V$};
\draw (626.98,283.94) node [anchor=north west][inner sep=0.75pt]    {$\phi $};
\draw (79.04,230.97) node [anchor=north west][inner sep=0.75pt]  [color={rgb, 255:red, 208; green, 2; blue, 27 }  ,opacity=1 ]  {$\phi _{0}$};
\draw (581,607.14) node [anchor=north west][inner sep=0.75pt]    {$x\rightarrow \infty $};
\draw (578.1,577.65) node [anchor=north west][inner sep=0.75pt]  [color={rgb, 255:red, 74; green, 144; blue, 226 }  ,opacity=1 ]  {$\phi \rightarrow \infty $};
\draw (97.93,572.19) node [anchor=north west][inner sep=0.75pt]  [color={rgb, 255:red, 208; green, 2; blue, 27 }  ,opacity=1 ]  {$\phi \rightarrow \phi _{0}$};
\draw (99.2,607.14) node [anchor=north west][inner sep=0.75pt]    {$x\rightarrow -\infty $};
\draw (292.16,570.05) node [anchor=north west][inner sep=0.75pt]   [align=left] {Brane solution};
\draw (138.56,404.75) node [anchor=north west][inner sep=0.75pt]   [align=left] {AdS throat};

\end{tikzpicture}
    \caption{Reconstruction of the brane (in red) from a scalar potential with a minimum at $\phi_0$ and $V \to 0$ as $\phi \to \infty$.}
    \label{AdStb}
\end{figure}

It is important to note that the limit $x \to -\infty$ corresponds to an infinite proper distance along the $x$-coordinate. However, this does not imply that it lies at infinite proper time for time-like observers. In fact, near $x \to -\infty$, the metric approaches the Poincaré patch of AdS, and thus the surface $x = -\infty$ represents a horizon that can be reached in finite proper time by infalling observers.

The solutions of the form~\eqref{DCm} and \eqref{DCs} in which a scalar field runs to an infinite distance in field space are known as \textit{Dynamical Cobordism} solutions\cite{Angius:2022aeq}. The original motivation for these solutions stems from the idea that, if the scalar field $\phi$ parametrizes a volume modulus and diverges at a finite proper distance in spacetime, then the internal compactification manifold collapses to zero volume. This realizes a defect where the extra dimensions shrink away, effectively terminating spacetime. Such defects are predicted by the Cobordism Conjecture to exist in a consistent theory of quantum gravity, as they trivialize the cobordism class of the internal manifold~\cite{McNamara:2019rup}.

However, whether such defects truly exist for manifolds with non-trivial cobordism classes is related to the UV completion of the theory. When the scalar field diverges at a finite distance, the effective field theory breaks down due to strong coupling or high curvature effects, rendering the classical description unreliable. In this work, we are instead interested in solutions where the scalar field runs to infinity only asymptotically. These configurations remain well within the regime of validity of the effective field theory and are therefore trustworthy.

It is worth noting that the idea of describing higher-dimensional brane solutions as lower-dimensional scalar field flows is not new. For example, many supersymmetric solutions were shown to arise as consistent truncations of higher-dimensional branes compactified on their transverse spheres~\cite{Angius:2022aeq}.

Most importantly, this perspective offers a profound physical insight: the black brane geometry that exhibits a given $\mathrm{AdS}_d$ space as its near-horizon geometry is, in the lower-dimensional theory, a scalar field solution that probes an infinite-distance limit of the moduli space. Thus, black brane solutions are inherently sensitive to the asymptotic behavior of the scalar potential far away in field space. We will make this connection more precise in Section~\ref{test}.

The idea of reconstructing branes from the lower-dimensional theory was also explored in~\cite{Apers:2025pon}, where the authors conjectured that the CFT dual of AdS always arises from stacks of branes probing a potentially singular geometry. Their method involves first removing fluxes to find the dynamical cobordism solution and then reinstating the branes dual to the fluxes. This construction differs from ours since our approach keeps fluxes throughout. However, in the regime where gravity is reliable, the two must agree once branes are added back. Importantly, our focus is on the asymptotic region far from the brane, which depends sensitively on the scalar potential in the opposite direction to that considered in~\cite{Apers:2025pon}.

Let us make an important remark. Even if the scalar potential admits a critical point and asymptotically vanishes in the infinite-distance region of field space, this does not automatically guarantee the existence of a solution of the form \eqref{DCm} and \eqref{DCs} that can be interpreted as an extremal black brane. If the portion of the scalar potential connecting the AdS critical point to the asymptotic region is monotonic, then the tools developed in the next subsection ensure that an extremal black brane solution always exists. However, if the scalar potential exhibits local maxima along the way, such a solution may fail to exist. It would therefore be very interesting to investigate in more detail which scalar potentials allow for extremal black brane solutions and which do not.

In the next subsection, we introduce a practical tool to develop intuition about these solutions. The key observation is that the geometry described by~\eqref{DCm} bears a striking resemblance to spatially flat FRW spacetimes sourced by scalar fields, with the important distinction that the spatial coordinate $x$ in our case plays the role typically associated with cosmological time. 

We will make this resemblance precise by demonstrating a one-to-one mapping between black brane solutions and Friedmann–Robertson–Walker (FRW) cosmologies, where the mapping inverts the sign of the scalar potential sourcing each solution. This mapping allows us to import results from cosmology—especially regarding the asymptotic behavior of solutions in scalar field space—directly into the context of black branes, without the need to rederive them from scratch. 

\subsection{Double Wick rotation}\label{Wick}

In this subsection, we study a helpful tool that allows us to quickly generate different black branes from the AdS solution. 

We highlight a precise correspondence between FRW cosmological solutions and the Dynamical Cobordism solutions, both characterized by a scale factor \( a \):

\begin{align}
\text{FRW:} \quad &ds^2 = -dt^2 + a(t)^2 \sum_{j=1}^{d} dx_j^2 \nonumber \\
\text{Dynamical cobordism:} \quad &ds^2 = a(x)^2 \left( -dt^2 + \sum_{i=1}^{d-2} d\sigma_i^2 \right) + dx^2
\end{align}

\noindent These solutions arise from a scalar field with potential \( V(\varphi) \), exhibiting high degrees of symmetry. The dynamical cobordism solution preserves Poincaré symmetry \( IO(1, d - 2) \), while the FRW background maintains symmetry \( IO(d - 1) \).

Upon solving Einstein’s equations, we identify the following dictionary:

\begin{align} \label{map}
t &\leftrightarrow x, \nonumber \\
a(t) &\leftrightarrow a(x), \nonumber \\
\varphi(t) &\leftrightarrow \varphi(x), \nonumber \\
V(\varphi) &\leftrightarrow -V(\varphi).
\end{align}

\noindent We refer to this mapping as a \emph{double Wick rotation}.

Starting with the dynamical cobordism metric, the relevant Christoffel symbols are:
\[
\Gamma^{\mu \neq x}_{\mu \neq x~x} = \Gamma^{\mu \neq x}_{x~\mu \neq x} = \frac{a'}{a}, \quad 
\Gamma^x_{ii} = -a a', \quad 
\Gamma^x_{tt} = a a'.
\]
where primes denote derivatives with respect to the coordinate \( x \). Here, \( \mu \neq x \) runs over all coordinates other than \( x \), including both spatial indices \( i \) and the temporal index \( t \). The non-zero Ricci tensor components are:
\[
R^\mu_\mu = -\frac{a''}{a} - (d - 2) \left( \frac{a'}{a} \right)^2, \quad 
R^x_x = -(d - 1) \frac{a''}{a}.
\]
The non-zero Einstein tensor components are:
\[
G^{\mu \neq x}_{\mu \neq x} = (d - 2)\frac{a''}{a} + \frac{(d - 2)(d - 3)}{2} \left( \frac{a'}{a} \right)^2, \quad
G^x_x = -\frac{(d - 1)(d - 2)}{2} \left( \frac{a'}{a} \right)^2.
\]

The energy-momentum tensor for a scalar field \( \varphi(x) \):
\[
T^\mu_\mu = -\frac{1}{2} (\varphi')^2 - V(\varphi), \quad 
T^x_x = \frac{1}{2} (\varphi')^2 - V(\varphi).
\]

Thus, the Einstein equations become:
\[
\frac{(d - 1)(d - 2)}{2} H^2 = \frac{1}{2} (\varphi')^2 - V(\varphi),
\]
\[
(d - 2) \frac{a''}{a} + \frac{(d - 2)(d - 3)}{2} H^2 = -\frac{1}{2} (\varphi')^2 - V(\varphi),
\]
where \( H = \frac{a'}{a} \).

Differentiating the first and manipulating appropriately gives the scalar field equation:
\[
\varphi'' + (d - 1) H \varphi' - \frac{dV}{d\varphi} = 0.
\]

For comparison, the FRW geometry has:
\[
\Gamma^j_{tj} = \Gamma^j_{jt} = \frac{\dot{a}}{a}, \quad \Gamma^t_{jj} = a \dot{a},
\]
\[
R^j_j = \frac{\ddot{a}}{a} + (d - 2)\left( \frac{\dot{a}}{a} \right)^2, \quad 
R^t_t = (d - 1)\frac{\ddot{a}}{a},
\]
\[
G^j_j = -(d - 2) \frac{\ddot{a}}{a} - \frac{(d - 2)(d - 3)}{2} H^2, \quad 
G^t_t = \frac{(d - 1)(d - 2)}{2} H^2,
\]
\[
T^j_j = \frac{1}{2} \dot{\varphi}^2 - V(\varphi), \quad 
T^t_t = -\frac{1}{2} \dot{\varphi}^2 - V(\varphi).
\]

Leading to the FRW equations:
\[
\frac{(d - 1)(d - 2)}{2} H^2 = \frac{1}{2} \dot{\varphi}^2 + V(\varphi),
\]
\[
\ddot{\varphi} + (d - 1) H \dot{\varphi} + \frac{dV}{d\varphi} = 0.
\]

These equations confirm the direct correspondence under the transformation \eqref{map}.
Using a double Wick rotation (DWR), one can map the black brane solution of~\eqref{DCm} with a negative scalar potential $V(\phi)$ to a spatially flat FRW solution with a positive scalar potential $V_{\mathrm{DWR}} = -V(\phi)$, which now has a maximum at $\phi_0$. The mapping is summarized as follows:
\begin{align}
\text{Black brane} &\longleftrightarrow \text{FRW} \nonumber\\
\lim_{x\rightarrow -\infty} \phi(x) \rightarrow \phi_0 &\longleftrightarrow \lim_{t\rightarrow -\infty} \phi(t) \rightarrow \phi_0 \nonumber\\
x \rightarrow -\infty: \quad a(x) \sim e^{x/L} &\longleftrightarrow t \rightarrow -\infty: \quad a(t) \sim e^{t/L} \nonumber\\
x \rightarrow \infty: \quad \phi(x) \rightarrow \infty &\longleftrightarrow t \rightarrow \infty: \quad \phi(t) \rightarrow \infty
\end{align}

Just as the black brane geometry develops an infinitely long throat in the limit $x \to -\infty$, the corresponding FRW solution features an infinitely extended time-like throat in the limit $t \to -\infty$. In this regime, the scale factor grows exponentially with time, implying that the geometry approaches a de Sitter spacetime in the distant past. This is consistent with the fact that the scalar field asymptotes to the maximum of the scalar potential at early times.

Therefore, the analog of Figure~\ref{AdStb} for the FRW solution obtained from the DWR of the black brane geometry is as follows:

\begin{figure}[H]
    \centering

  
\tikzset {_ayj9ncwej/.code = {\pgfsetadditionalshadetransform{ \pgftransformshift{\pgfpoint{0 bp } { 0 bp }  }  \pgftransformrotate{0 }  \pgftransformscale{2 }  }}}
\pgfdeclarehorizontalshading{_i03jlugws}{150bp}{rgb(0bp)=(1,1,0);
rgb(37.5bp)=(1,1,0);
rgb(62.5bp)=(0,0.5,0.5);
rgb(100bp)=(0,0.5,0.5)}
\tikzset{every picture/.style={line width=0.75pt}} 

\begin{tikzpicture}[x=0.75pt,y=0.75pt,yscale=-1,xscale=1]

\draw  (3.35,257.26) -- (637.76,257.26)(63.58,24.9) -- (63.58,367.26) (630.76,252.26) -- (637.76,257.26) -- (630.76,262.26) (58.58,31.9) -- (63.58,24.9) -- (68.58,31.9)  ;
\draw    (-13.5,240) .. controls (22.5,209) and (13.5,144) .. (80.5,131) .. controls (147.5,118) and (263.5,246) .. (605.5,250) ;
\draw [color={rgb, 255:red, 208; green, 2; blue, 27 }  ,draw opacity=1 ] [dash pattern={on 4.5pt off 4.5pt}]  (83.5,258) -- (84.14,131.7) ;
\draw [shading=_i03jlugws,_ayj9ncwej][line width=1.5]    (98.97,589.7) -- (627.52,589.7) ;
\draw [shift={(630.52,589.7)}, rotate = 180] [color={rgb, 255:red, 0; green, 0; blue, 0 }  ][line width=1.5]    (14.21,-4.28) .. controls (9.04,-1.82) and (4.3,-0.39) .. (0,0) .. controls (4.3,0.39) and (9.04,1.82) .. (14.21,4.28)   ;
\draw    (92.24,423) .. controls (233.73,419.94) and (451.78,455.68) .. (620.22,316.29) ;
\draw    (91.49,453) .. controls (209.87,451.21) and (522.9,435.13) .. (618.35,538.77) ;
\draw  [fill={rgb, 255:red, 74; green, 144; blue, 226 }  ,fill opacity=1 ] (597.31,477.84) .. controls (590.33,422.27) and (596.59,354.77) .. (611.28,327.07) .. controls (625.97,299.37) and (643.53,321.97) .. (650.5,377.54) .. controls (657.47,433.11) and (651.21,500.61) .. (636.52,528.31) .. controls (621.84,556) and (604.28,533.41) .. (597.31,477.84) -- cycle ;
\draw  [fill={rgb, 255:red, 208; green, 2; blue, 27 }  ,fill opacity=1 ] (88.52,444.76) .. controls (87.59,437.39) and (88.42,428.45) .. (90.37,424.78) .. controls (92.31,421.12) and (94.64,424.11) .. (95.56,431.47) .. controls (96.49,438.83) and (95.66,447.77) .. (93.71,451.44) .. controls (91.77,455.11) and (89.44,452.12) .. (88.52,444.76) -- cycle ;
\draw    (260.4,411.9) -- (97.23,411.01) ;
\draw [shift={(95.23,411)}, rotate = 0.31] [color={rgb, 255:red, 0; green, 0; blue, 0 }  ][line width=0.75]    (10.93,-3.29) .. controls (6.95,-1.4) and (3.31,-0.3) .. (0,0) .. controls (3.31,0.3) and (6.95,1.4) .. (10.93,3.29)   ;

\draw (40.54,16.43) node [anchor=north west][inner sep=0.75pt]    {$V$};
\draw (622.82,273.2) node [anchor=north west][inner sep=0.75pt]    {$\phi $};
\draw (73.89,266.23) node [anchor=north west][inner sep=0.75pt]  [color={rgb, 255:red, 208; green, 2; blue, 27 }  ,opacity=1 ]  {$\phi _{0}$};
\draw (578.85,596.4) node [anchor=north west][inner sep=0.75pt]    {$t\rightarrow \infty $};
\draw (575.94,566.91) node [anchor=north west][inner sep=0.75pt]  [color={rgb, 255:red, 74; green, 144; blue, 226 }  ,opacity=1 ]  {$\phi \rightarrow \infty $};
\draw (95.77,561.45) node [anchor=north west][inner sep=0.75pt]  [color={rgb, 255:red, 208; green, 2; blue, 27 }  ,opacity=1 ]  {$\phi \rightarrow \phi _{0}$};
\draw (97.05,596.4) node [anchor=north west][inner sep=0.75pt]    {$t\rightarrow -\infty $};
\draw (290.01,559.31) node [anchor=north west][inner sep=0.75pt]   [align=left] {S-Brane solution};
\draw (136.4,394.01) node [anchor=north west][inner sep=0.75pt]   [align=left] {dS throat};

\end{tikzpicture}
    \caption{The Double Wick Rotation of the black brane solution is an FRW solution which in the past infinity asymptotes to de Sitter space and in the future infinity is a scalar field cosmology driven by an exponential positive scalar potential.}
    \label{dSt}
\end{figure}
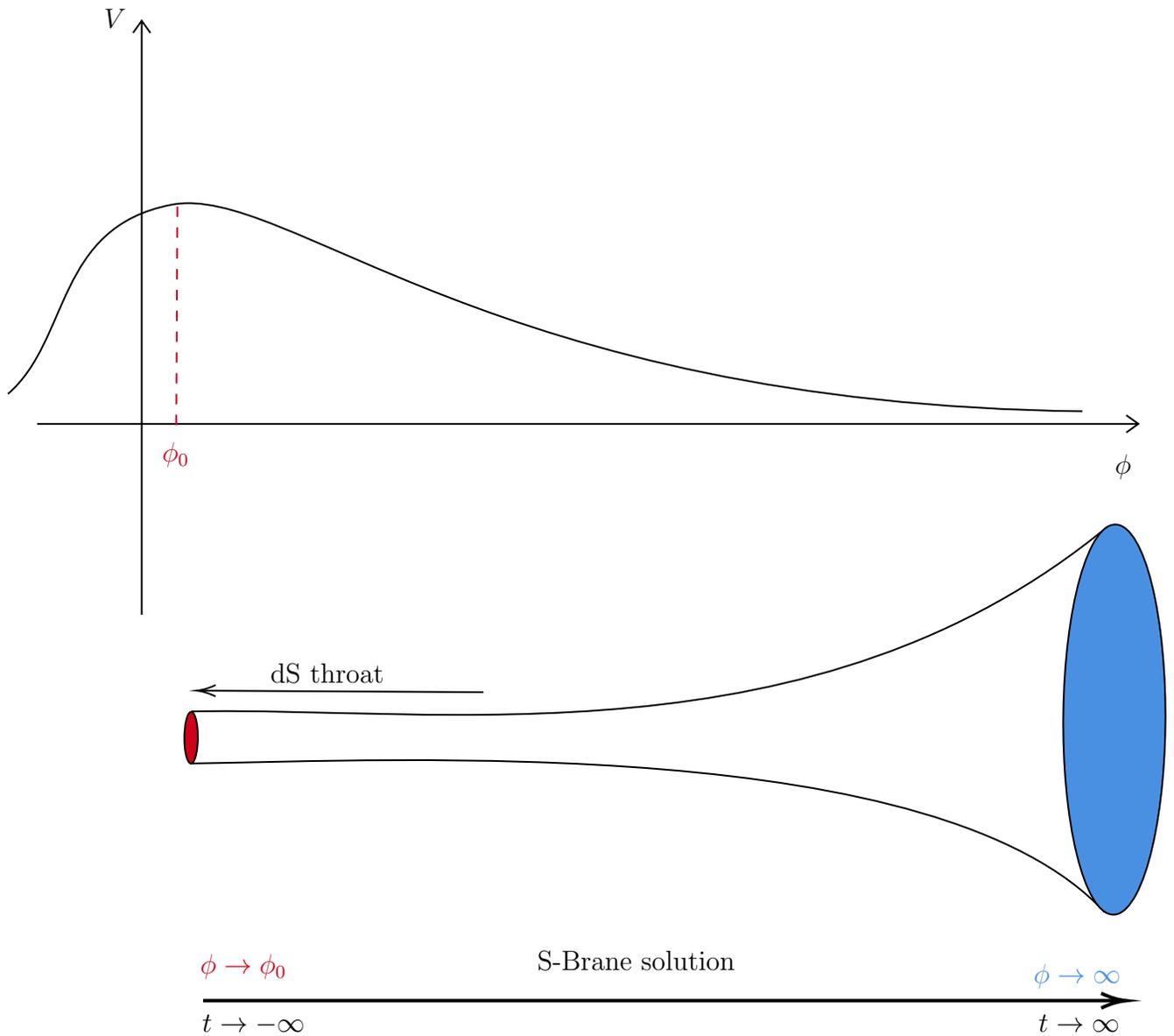

In the black brane solution, the surface at $x \rightarrow -\infty$ lies at an infinite spatial proper distance. However, since it corresponds to the Poincaré horizon of the near-horizon AdS geometry, it remains traversable by timelike observers. Similarly, in the FRW solution, the limit $t \rightarrow -\infty$ corresponds to a null surface associated with the past infinity in the flat slicing of de Sitter space, which is also accessible to timelike observers.

We now study the behavior of the black brane solution far from the brane, i.e., the $x \rightarrow \infty$ limit. Under the double Wick rotation, this maps to the $t \rightarrow \infty$ behavior of the FRW solution, which is driven by an exponential scalar potential. Such FRW cosmologies have been extensively studied. Below, we review the known classification of FRW attractor solutions driven by exponential potentials and then use the DWR to infer the behavior of the corresponding black brane solutions at large $x$.

In the FRW case, scalar cosmologies with exponential potentials admit attractor solutions whose asymptotic behavior depends on the steepness of the potential. A critical slope separates qualitatively different classes of solutions. Specifically, the threshold is given by
\begin{align}\label{threshold}
\lambda=2\sqrt{\frac{d-1}{d-2}} \,,
\end{align}
where $\lambda=\left| \frac{\nabla V}{V} \right|$. If the potential is steeper than this threshold, the scalar potential falls off rapidly and becomes negligible compared to the kinetic energy of the scalar field. These solutions are known as \emph{kination} solutions. In this regime, regardless of the precise steepness of the exponential, the scale factor and scalar field evolve asymptotically as:
\begin{align}\label{AB1}
    a(t) &\sim t^{\frac{1}{d-1}}, \\
    \phi(t) &\simeq \sqrt\frac{d-2}{d-1}\log t\,.
\end{align}

On the other hand, if the slope of the potential is below the threshold \eqref{threshold}, the fractional contribution of the scalar potential to the energy density asymptotically converges to a non-zero constant. In this case, the cosmological attractor solution is given by
\begin{align}\label{AB2}
    a(t) &\sim t^{\frac{4}{(d-2)\lambda^2}}, \\
    \phi(t) &\simeq \frac{2}{\lambda}\log t\,.
\end{align}

Using double Wick rotation \eqref{map}, we can map these behaviors to the asymptotic profile of the black brane solutions. We find that depending on the value of $\lambda=|\frac{\nabla V}{V}|$, there are two qualitatively distinct asymptotic behaviors:
\begin{align}\label{AB}
    &\lambda\geq 2\sqrt\frac{d-1}{d-2}:     a(x) \sim x^{\frac{1}{d-1}},~~\phi(x) \simeq \sqrt\frac{d-2}{d-1}\log x\nonumber\\
    &\lambda< 2\sqrt\frac{d-1}{d-2}:     a(x) \sim x^{\frac{4}{(d-2)\lambda^2}},~~\phi(x) \simeq \frac{2}{\lambda}\log x\,.
\end{align}

In the presence of multiple scalar fields, the potential $V(\vec{\phi})$ can have different slopes in different directions. However, the relevant direction for the black brane solution will correspond to the gradient direction. This becomes clear when we consider the corresponding FRW solution obtained via double Wick rotation: in cosmological evolution, scalar fields rolling on an exponential potential will asymptotically align along the direction of the steepest slope, even if their initial velocity $\dot{\vec{\phi}}$ pointed elsewhere. This alignment ensures that the effective dynamics are governed by the dominant exponential term, reducing the system to an effectively single-field behavior at late times (or far distances in the black brane geometry). This technical point underlies our formulation of the Asymptotic No-Scale Separation condition in Section~\ref{ANSS}, where we defined infinite distance limits to specifically refer to directions in scalar field space along which the scalar potential \( V \to 0 \).

\subsection{Non-uniqueness of the solution}

At the beginning of this section, we showed that a black brane solution which develops a given AdS near-horizon geometry arises as a solution to the scalar field equations, where the scalar field depends only on a single coordinate. Such a solution asymptotes to the AdS value in one spacetime direction while diverging in the opposite direction. In the previous subsection, we demonstrated that these solutions can be obtained from spatially flat FRW solutions for the scalar potential $-V$. 

We now highlight a nontrivial corollary of our findings: there can be multiple solutions with the same AdS near-horizon geometry. These distinct solutions correspond to scalar fields evolving along different directions of the moduli space.

\begin{figure}[H]
    \centering

\tikzset{every picture/.style={line width=0.75pt}} 

\begin{tikzpicture}[x=0.75pt,y=0.75pt,yscale=-1,xscale=1]

\draw   (324,221.64) -- (407.93,305.57) -- (324,389.5) -- (240.07,305.57) -- cycle ;
\draw   (324,53.79) -- (407.93,137.72) -- (324,221.64) -- (240.07,137.72) -- cycle ;
\draw    (240.5,9) .. controls (242.17,10.67) and (242.17,12.33) .. (240.51,14) .. controls (238.85,15.67) and (238.85,17.33) .. (240.52,19) .. controls (242.19,20.66) and (242.2,22.33) .. (240.54,24) .. controls (238.88,25.67) and (238.88,27.33) .. (240.55,29) .. controls (242.22,30.67) and (242.22,32.33) .. (240.56,34) .. controls (238.9,35.67) and (238.9,37.33) .. (240.57,39) .. controls (242.24,40.66) and (242.25,42.33) .. (240.59,44) .. controls (238.93,45.67) and (238.93,47.33) .. (240.6,49) .. controls (242.27,50.67) and (242.27,52.33) .. (240.61,54) .. controls (238.95,55.67) and (238.95,57.33) .. (240.62,59) .. controls (242.29,60.67) and (242.29,62.33) .. (240.63,64) .. controls (238.97,65.67) and (238.98,67.34) .. (240.65,69) .. controls (242.32,70.67) and (242.32,72.33) .. (240.66,74) .. controls (239,75.67) and (239,77.33) .. (240.67,79) .. controls (242.34,80.67) and (242.34,82.33) .. (240.68,84) .. controls (239.02,85.67) and (239.03,87.34) .. (240.7,89) .. controls (242.37,90.67) and (242.37,92.33) .. (240.71,94) .. controls (239.05,95.67) and (239.05,97.33) .. (240.72,99) .. controls (242.39,100.67) and (242.39,102.33) .. (240.73,104) .. controls (239.07,105.67) and (239.07,107.33) .. (240.74,109) .. controls (242.41,110.66) and (242.42,112.33) .. (240.76,114) .. controls (239.1,115.67) and (239.1,117.33) .. (240.77,119) .. controls (242.44,120.67) and (242.44,122.33) .. (240.78,124) .. controls (239.12,125.67) and (239.12,127.33) .. (240.79,129) .. controls (242.46,130.66) and (242.47,132.33) .. (240.81,134) .. controls (239.15,135.67) and (239.15,137.33) .. (240.82,139) .. controls (242.49,140.67) and (242.49,142.33) .. (240.83,144) .. controls (239.17,145.67) and (239.17,147.33) .. (240.84,149) .. controls (242.51,150.67) and (242.51,152.33) .. (240.85,154) .. controls (239.19,155.67) and (239.2,157.34) .. (240.87,159) .. controls (242.54,160.67) and (242.54,162.33) .. (240.88,164) .. controls (239.22,165.67) and (239.22,167.33) .. (240.89,169) .. controls (242.56,170.67) and (242.56,172.33) .. (240.9,174) .. controls (239.24,175.67) and (239.25,177.34) .. (240.92,179) .. controls (242.59,180.67) and (242.59,182.33) .. (240.93,184) .. controls (239.27,185.67) and (239.27,187.33) .. (240.94,189) .. controls (242.61,190.67) and (242.61,192.33) .. (240.95,194) .. controls (239.29,195.67) and (239.29,197.33) .. (240.96,199) .. controls (242.63,200.66) and (242.64,202.33) .. (240.98,204) .. controls (239.32,205.67) and (239.32,207.33) .. (240.99,209) .. controls (242.66,210.67) and (242.66,212.33) .. (241,214) .. controls (239.34,215.67) and (239.34,217.33) .. (241.01,219) .. controls (242.68,220.66) and (242.69,222.33) .. (241.03,224) .. controls (239.37,225.67) and (239.37,227.33) .. (241.04,229) .. controls (242.71,230.67) and (242.71,232.33) .. (241.05,234) .. controls (239.39,235.67) and (239.39,237.33) .. (241.06,239) .. controls (242.73,240.67) and (242.73,242.33) .. (241.07,244) .. controls (239.41,245.67) and (239.42,247.34) .. (241.09,249) .. controls (242.76,250.67) and (242.76,252.33) .. (241.1,254) .. controls (239.44,255.67) and (239.44,257.33) .. (241.11,259) .. controls (242.78,260.67) and (242.78,262.33) .. (241.12,264) .. controls (239.46,265.67) and (239.47,267.34) .. (241.14,269) .. controls (242.81,270.67) and (242.81,272.33) .. (241.15,274) .. controls (239.49,275.67) and (239.49,277.33) .. (241.16,279) .. controls (242.83,280.67) and (242.83,282.33) .. (241.17,284) .. controls (239.51,285.67) and (239.51,287.33) .. (241.18,289) .. controls (242.85,290.66) and (242.86,292.33) .. (241.2,294) .. controls (239.54,295.67) and (239.54,297.33) .. (241.21,299) .. controls (242.88,300.67) and (242.88,302.33) .. (241.22,304) .. controls (239.56,305.67) and (239.56,307.33) .. (241.23,309) .. controls (242.9,310.66) and (242.91,312.33) .. (241.25,314) .. controls (239.59,315.67) and (239.59,317.33) .. (241.26,319) .. controls (242.93,320.67) and (242.93,322.33) .. (241.27,324) .. controls (239.61,325.67) and (239.61,327.33) .. (241.28,329) .. controls (242.95,330.67) and (242.95,332.33) .. (241.29,334) .. controls (239.63,335.67) and (239.64,337.34) .. (241.31,339) .. controls (242.98,340.67) and (242.98,342.33) .. (241.32,344) .. controls (239.66,345.67) and (239.66,347.33) .. (241.33,349) .. controls (243,350.67) and (243,352.33) .. (241.34,354) .. controls (239.68,355.67) and (239.69,357.34) .. (241.36,359) .. controls (243.03,360.67) and (243.03,362.33) .. (241.37,364) .. controls (239.71,365.67) and (239.71,367.33) .. (241.38,369) .. controls (243.05,370.67) and (243.05,372.33) .. (241.39,374) .. controls (239.73,375.67) and (239.73,377.33) .. (241.4,379) .. controls (243.07,380.66) and (243.08,382.33) .. (241.42,384) .. controls (239.76,385.67) and (239.76,387.33) .. (241.43,389) .. controls (243.1,390.67) and (243.1,392.33) .. (241.44,394) .. controls (239.78,395.67) and (239.78,397.33) .. (241.45,399) .. controls (243.12,400.66) and (243.13,402.33) .. (241.47,404) .. controls (239.81,405.67) and (239.81,407.33) .. (241.48,409) .. controls (243.15,410.67) and (243.15,412.33) .. (241.49,414) -- (241.5,418) -- (241.5,418) ;

\draw (320,21.4) node [anchor=north west][inner sep=0.75pt]    {$\vdots $};
\draw (319,395.4) node [anchor=north west][inner sep=0.75pt]    {$\vdots $};
\draw (243,213) node [anchor=north west][inner sep=0.75pt]  [font=\small] [align=left] {Solution II};
\draw (298,300) node [anchor=north west][inner sep=0.75pt]  [font=\small] [align=left] {Solution I};
\draw (294,120) node [anchor=north west][inner sep=0.75pt]  [font=\small] [align=left] {Solution I};
\draw (214.9,234.5) node [anchor=north west][inner sep=0.75pt]  [rotate=-270]  {$R=0$};
\draw (275.43,141.77) node [anchor=north west][inner sep=0.75pt]  [rotate=-45]  {$R=R_{H}$};
\draw (267.27,282.07) node [anchor=north west][inner sep=0.75pt]  [rotate=-315]  {$R=R_{H}$};

\end{tikzpicture}
\caption{Maximally extended Penrose diagram of the extremal 3-brane in 10-dimensional type IIB supergravity. Both solutions~I and~II are described by scalar field configurations arising in the 5D compactification of type IIB on $S^5$, as discussed in Subsection~\ref{DC}. In both cases, the volume modulus is fixed to its critical value along the null surface, and departs from this value as one moves away from the surface. In Solution~I, the size of the $S^5$ grows without bound at infinite spatial distance, whereas in Solution~II the size of the $S^5$ shrinks to zero at finite spatial distance.}

    \label{NU}
\end{figure}
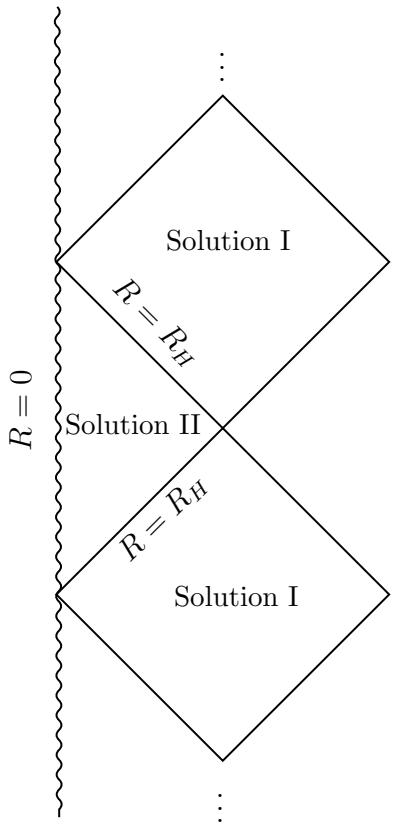

Consider the example of $\mathrm{AdS}_5 \times S^5$. In one black brane solution, the scalar field runs along the decompactification direction. Alternatively, the scalar field may roll in a direction where the $S^5$ shrinks to zero size. In the latter case, the presence of fluxes ensures that the geometry develops a singularity at finite distance. Remarkably, the two solutions are not different and belong to the same spacetime as shown in Figure~\ref{NU}. One solution describes the exterior of the black brane while the other one describes the interior of the black brane. 

Note that the solution corresponding to the interior of the black brane does not play a role in the holographic argument for the duality between the worldvolume theory of D3-branes and \( \mathrm{AdS}_5 \). As we will make more precise in the next section, a key ingredient in holography is the existence of a decoupling limit. This requires not only that the spacetime possess an asymptotic spatial infinity away from the black brane, but also that this region satisfies a suitable decoupling condition.

In the discussion above, we illustrated that the asymptotic behavior of black brane solutions depends on the direction in which the scalar field rolls in field space. However, it is important to emphasize that the number of qualitatively distinct solutions is finite. 

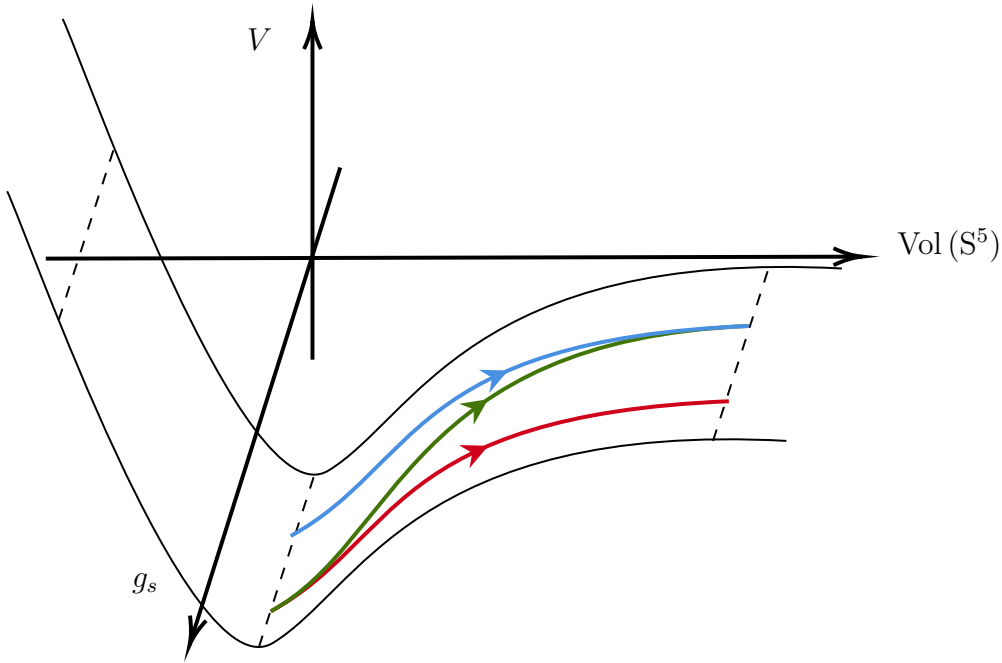
\begin{figure}[H]
    \centering

\tikzset{every picture/.style={line width=0.75pt}} 

\begin{tikzpicture}[x=0.75pt,y=0.75pt,yscale=-1,xscale=1]

\draw    (94.5,110) .. controls (105.5,133) and (185.5,364) .. (228.5,338) .. controls (271.5,312) and (300.5,226) .. (487.5,236) ;
\draw    (122.5,23) .. controls (133.5,46) and (213.5,277) .. (256.5,251) .. controls (299.5,225) and (328.5,139) .. (515.5,149) ;
\draw  [dash pattern={on 4.5pt off 4.5pt}]  (221.5,340) -- (249.5,253) ;
\draw  [dash pattern={on 4.5pt off 4.5pt}]  (450.5,236) -- (478.5,149) ;
\draw  [dash pattern={on 4.5pt off 4.5pt}]  (120.5,175) -- (148.5,88) ;
\draw [color={rgb, 255:red, 208; green, 2; blue, 27 }  ,draw opacity=1 ][line width=1.5]    (227.5,322) .. controls (291.5,287) and (286.5,223) .. (458.5,216) ;
\draw [shift={(337.21,238.46)}, rotate = 155.7] [fill={rgb, 255:red, 208; green, 2; blue, 27 }  ,fill opacity=1 ][line width=0.08]  [draw opacity=0] (13.4,-6.43) -- (0,0) -- (13.4,6.44) -- (8.9,0) -- cycle    ;
\draw [color={rgb, 255:red, 65; green, 117; blue, 5 }  ,draw opacity=1 ][line width=1.5]    (227.5,322) .. controls (291.5,287) and (296.5,185) .. (468.5,178) ;
\draw [shift={(337.13,214.94)}, rotate = 145.22] [fill={rgb, 255:red, 65; green, 117; blue, 5 }  ,fill opacity=1 ][line width=0.08]  [draw opacity=0] (13.4,-6.43) -- (0,0) -- (13.4,6.44) -- (8.9,0) -- cycle    ;
\draw [color={rgb, 255:red, 74; green, 144; blue, 226 }  ,draw opacity=1 ][line width=1.5]    (237.5,284) .. controls (301.5,249) and (296.5,185) .. (468.5,178) ;
\draw [shift={(347.21,200.46)}, rotate = 155.7] [fill={rgb, 255:red, 74; green, 144; blue, 226 }  ,fill opacity=1 ][line width=0.08]  [draw opacity=0] (13.4,-6.43) -- (0,0) -- (13.4,6.44) -- (8.9,0) -- cycle    ;
\draw [line width=1.5]    (262.5,98) -- (187.4,336.14) ;
\draw [shift={(186.5,339)}, rotate = 287.5] [color={rgb, 255:red, 0; green, 0; blue, 0 }  ][line width=1.5]    (14.21,-4.28) .. controls (9.04,-1.82) and (4.3,-0.39) .. (0,0) .. controls (4.3,0.39) and (9.04,1.82) .. (14.21,4.28)   ;
\draw [line width=1.5]    (114,144) -- (522.5,143.01) ;
\draw [shift={(525.5,143)}, rotate = 179.86] [color={rgb, 255:red, 0; green, 0; blue, 0 }  ][line width=1.5]    (14.21,-4.28) .. controls (9.04,-1.82) and (4.3,-0.39) .. (0,0) .. controls (4.3,0.39) and (9.04,1.82) .. (14.21,4.28)   ;
\draw [line width=1.5]    (248.5,195) -- (248.5,27) ;
\draw [shift={(248.5,24)}, rotate = 90] [color={rgb, 255:red, 0; green, 0; blue, 0 }  ][line width=1.5]    (14.21,-4.28) .. controls (9.04,-1.82) and (4.3,-0.39) .. (0,0) .. controls (4.3,0.39) and (9.04,1.82) .. (14.21,4.28)   ;

\draw (156,301.4) node [anchor=north west][inner sep=0.75pt]    {$g_{s}$};
\draw (542,127.4) node [anchor=north west][inner sep=0.75pt]    {$\rm{Vol}\left( S^{5}\right)$};
\draw (214,26.4) node [anchor=north west][inner sep=0.75pt]    {$V$};

\end{tikzpicture}
    \caption{Scalar potential as a function of the string coupling \( g_s \) and the volume modulus of \( S^5 \) in type IIB on $S^5$ with \( N \) flux units. Red, green, and blue curves denote scalar field trajectories that uplift to black brane solutions. The asymptotic value of \( g_s \) fixes the ten-dimensional background moduli, while its initial value determines the \( \mathrm{AdS}_5 \) modulus in the near-horizon geometry. The red and green curves share the same near-horizon coupling but differ asymptotically, whereas the green and blue curves share the same asymptotic coupling but differ near the horizon.}
    \label{NUP}
\end{figure}

For instance, in the case of the \( \mathrm{AdS}_5 \times S^5 \) background, one can consider scalar field velocities \( \dot{\phi} \) that point in a direction that is a linear combination of the volume modulus of the internal \( S^5 \) and variations of the dilaton. However, since the scalar potential does not depend on the dilaton, the gradient flow aligns the trajectory in field space along the direction of increasing volume modulus of \( S^5 \), while the dilaton asymptotes to a constant value. Consequently, different initial conditions for the dilaton variation lead to the same class of extremal black brane solutions in ten-dimensional Minkowski space, but with different asymptotic values of the string coupling \( g_s \). These solutions correspond to the same infinite-distance limit in field space but differ by parallel trajectories labeled by the asymptotic value of the frozen dilaton (Figure~\ref{NUP}).

This residual freedom reflects the moduli of the underlying string background. In particular, the dilaton modulus parametrizes the string coupling in both the asymptotic flat background and the \( \mathrm{AdS}_5 \) near-horizon geometry. Based on the finiteness of black hole entropy, it was argued in~\cite{Bedroya:2023tch} that the number of inequivalent infinite-distance limits is finite modulo self-dualities. Therefore, while each infinite-distance direction may admit a continuous family of solutions labeled by frozen moduli, the number of qualitatively distinct black brane solutions is finite.

\subsection{Examples}

Let us revisit the examples of extremal branes discussed in Section~\ref{Ext}, and reinterpret these black branes as scalar field solutions in a lower-dimensional theory obtained by truncating the higher-dimensional equations of motion. We considered four broad classes of solutions:
\begin{itemize}
    \item Regular extremal branes in Minkowski space,
    \item Extremal branes with singular horizons,
    \item Extremal branes with non-Minkowski Ricci-flat asymptotics,
    \item Extremal branes in AdS spacetimes.
\end{itemize}

\paragraph{Regular extremal branes in Minkowski space:}
Consider an extremal \( p \)-brane in \( D \)-dimensional Minkowski spacetime. Its metric takes the form:
\begin{align}
    ds^2 &\;\simeq\; f(\rho)^2\left[-dt^2 + d\vec{y}_p^2\right] + d\rho^2 + R(\rho)^2\, d\Omega_{D-p-2}^2 \\
    \rho \rightarrow -\infty &: \quad f(\rho)\sim e^{\rho/L}, \quad R(\rho) \to \text{const.} \\
    \rho \rightarrow \infty &: \quad f(\rho)\to 1, \quad R(\rho)\simeq \rho.
\end{align}

As discussed in Section~\ref{MinB}, the near-horizon geometry is \( \mathrm{AdS}_{p+1} \times S^{D-p-2} \), where the transverse sphere links with the brane. This AdS region can be derived via a Freund--Rubin compactification~\cite{Freund:1980xh}, obtained by reducing the \( D \)-dimensional theory on \( S^{D-p-2} \) with fluxes sourced by the brane threading the compact sphere. 

From the perspective of the lower-dimensional effective theory (with dimension \( p+2 \)), the resulting black \( p \)-brane corresponds to a dynamical cobordism solution of the type:
\begin{align}
    ds^2 &\;\simeq\; a(x)^2\left[-dt^2 + d\vec{y}_p^2\right] + dx^2 \\
    x \rightarrow -\infty &: \quad a(x)\sim e^{x/L} \\
    x \rightarrow \infty &: \quad a(x)\sim x^{\frac{D-p-2}{D-2}}.
\end{align}

It is important to emphasize that the coordinates \( \rho \) and \( x \) are not identical since they are proper distances measured with respect to different metrics. The two metrics (higher- and lower-dimensional Einstein frames) are related via a Weyl rescaling. As a result, the functions \( f(\rho) \) and \( a(x) \) are also not identical, though they encode the same geometric structure in different dimensions.

Note that the asymptotic behavior of \( a(x) \) as \( x \rightarrow \infty \) is consistent with the result in \eqref{AB} for 
\[
\left|\frac{\nabla V}{V}\right| = 2\sqrt{\frac{D - 2}{(D - d)(d - 2)}},
\]
which corresponds to the falloff rate associated with the curvature of a \( (D - d) \)-dimensional internal Einstein space.
This construction generalizes the uplift of the 3-brane in the familiar \( \mathrm{AdS}_5 \times S^5 \) background, which we analyzed in Section~\ref{uplift}.

\paragraph{Singular extremal branes in Minkowski space:}
In the case of singular branes, there is no AdS near-horizon geometry, reflecting the absence of a minimum in the scalar potential obtained via compactification on the transverse sphere. Nevertheless, one can still construct dynamical cobordism solutions by truncating the higher-dimensional theory on the transverse sphere and interpreting the singular brane as a solution in the lower-dimensional effective theory.

A classic example is a stack of NS5-brane, which can be viewed as a 7-dimensional solution obtained by compactifying type II string theory on an \( S^3 \) threaded by NS-NS 3-form \( H \)-flux. The resulting scalar potential lacks a minimum due to the dilaton dependence of the gauge coupling. Consequently, the solution is singular at a finite value of the proper distance coordinate \( x \). 

Despite this singularity, the conclusions drawn in Section~\ref{Wick} regarding the behavior of solutions far from the brane remain valid. This is because those conclusions depend only on the asymptotic behavior of the scalar potential in the infinite distance regime, where the effective field theory remains trustworthy.

For a stack of NS5-branes, as discussed in \eqref{NS5}, the higher-dimensional metric takes the form:
\begin{align}
    ds^2 &\;\simeq\; f(\rho)^2\left[-dt^2 + {d\vec{y}_5}^2\right] + d\rho^2 + R(\rho)^2\, d\Omega_{3}^2, \\
    \rho \rightarrow 0 &: \quad f(\rho)\sim \left(\frac{\rho}{L}\right)^2, \quad R(\rho)\sim \left(\frac{\rho}{L}\right)^6, \\
    \rho \rightarrow \infty &: \quad f(\rho)\to 1, \quad R(\rho)\sim \rho.
\end{align}

The near-horizon geometry is a linear dilaton background, while far away from the brane the spacetime asymptotes to 10-dimensional Minkowski space. From the perspective of the lower-dimensional effective theory obtained by compactifying the 10d theory on \( S^3 \), the NS5-brane corresponds to a dynamical cobordism solution of the form:
\begin{align}
    ds^2 &\;\simeq\; a(x)^2\left[-dt^2 + {d\vec{y}_5}^2\right] + dx^2, \\
    x \rightarrow 0 &: \quad a(x)\sim \left(\frac{x}{L}\right)^{\frac{4}{11}}, \\
    x \rightarrow \infty &: \quad a(x)\sim x^{\frac{3}{8}}.
\end{align}

Note that the asymptotic behavior of \( a(x) \) as \( x \rightarrow \infty \) is consistent with the result in \eqref{AB} for 
\[
\left|\frac{\nabla V}{V}\right| = 2\sqrt{\frac{8}{15}},
\]
which corresponds to the falloff rate associated with the contribution of the curvature of $S^3$ to the scalar potential.
Note that the behavior as \( x \rightarrow \infty \) matches that of regular branes, since both solutions asymptote to Minkowski spacetime in the far region.

\paragraph{Extremal branes with Ricci-flat and non-Minkowski asymptotics:}

Next, consider the case of regular branes with Ricci-flat but non-globally Minkowski asymptotics. For example, in Section~\ref{exotic}, we reviewed a solution that realizes the Klebanov--Witten AdS vacuum in its near-horizon geometry. This geometry is structurally similar to \( \mathrm{AdS}_5 \times S^5 \), except that the internal \( S^5 \) is replaced with a five-dimensional Einstein manifold \( X \). Crucially, the 5D scalar potential, viewed as a function of the volume modulus, is insensitive to the detailed shape of the internal manifold. Consequently, the black brane solution from the 5D perspective remains unchanged.

The 10D metric for the extremal black branes that realize such AdS spaces takes the form:
\begin{align}
    ds^2 = f(\rho)^2 \left[-dt^2 + d\vec{y}_p^2\right] + d\rho^2 + R(\rho)^2\, ds_X^2,
\end{align}
where \( f(\rho) \) is identical to that of the black 3-brane in type IIB string theory with \( \mathrm{AdS}_5 \times S^5 \) near-horizon geometry. Therefore, when we truncate the equations of motion on the compact five-manifold (in this case \( X \)), the resulting 5D metric takes the same form:
\begin{align}
    ds^2 &\;\simeq\; a(x)^2\left[-dt^2 + d\vec{y}_3^2\right] + dx^2, \\
    x \rightarrow -\infty &: \quad a(x)\sim e^{x/L}, \\
    x \rightarrow \infty &: \quad a(x)\sim x^{\frac{5}{8}}.
\end{align}
Note that the asymptotic behavior of \( a(x) \) as \( x \rightarrow \infty \) is consistent with the result in \eqref{AB} for 
\[
\left|\frac{\nabla V}{V}\right| = 2\sqrt{\frac{8}{15}},
\]
which corresponds to the falloff rate associated with the contribution of the curvature of $S^5$ to the scalar potential.

\paragraph{Extremal branes in AdS spacetimes:}

Finally, let us consider the most non-trivial case: extremal black branes embedded in ambient AdS spacetimes. These solutions, reviewed in Section~\ref{AdSdS}, interpolate between two AdS regions of different dimensionality: one appearing in the near-horizon region, and the other in the asymptotic geometry far from the brane. In particular, the example discussed in Section~\ref{AdSdS} has near-horizon geometry \( \mathrm{AdS}_3 \times \Sigma \), where \( \Sigma \) is a Riemann surface, while the asymptotic geometry is \( \mathrm{AdS}_5 \).

To analyze this configuration from the perspective of a 3D effective theory, we compactify the full 5D solution \eqref{5DRS} on the Riemann surface \( \Sigma \). The 5D metric takes the form:
\begin{align}
ds_5^2 &= f(\rho)^2(-dt^2 + dz^2) + d\rho^2 + e^{2g(\rho) + 2h(x,y)}(dx^2 + dy^2), \\
\rho \rightarrow -\infty &: \quad f(\rho) \sim e^{\rho / L_{\mathrm{AdS}_3}}, \quad g(\rho) \rightarrow \text{const}, \\
\rho \rightarrow \infty &: \quad f(\rho) \sim e^{\rho / L_{\mathrm{AdS}_5}}, \quad g(\rho) \sim \frac{\rho}{L_{\mathrm{AdS}_5}}.
\end{align}

Upon dimensional reduction on \( \Sigma \), the 3D Einstein-frame metric becomes:
\begin{align}
ds_3^2 &= a(x)^2(-dt^2 + dz^2) + dx^2, \\
x \rightarrow -\infty &: \quad a(x) \sim e^{x / L_{\mathrm{AdS}_3}}, \\
x \rightarrow \infty &: \quad a(x) \sim x^{\frac{3}{2}}.
\end{align}

There are three scalar fields in the effective theory: one arising from the volume modulus of \( \Sigma \), and two others associated with squashing modes of the \( S^5 \) fiber in the 10D type IIB background (see Section~\ref{AdSdS} for their origin). Among these, the squashing modes approach constant values as \( x \rightarrow \infty \), while the volume modulus of \( \Sigma \) diverges.

In the 3D Einstein frame, the scalar potential receives contributions from the 5D cosmological constant and the curvature of the internal Riemann surface. Both contributions decay exponentially as the volume of \( \Sigma \) grows. However, the contribution from the 5D cosmological constant dominates asymptotically and governs the large-\( x \) behavior of the solution. 

Notably, the asymptotic scaling \( a(x) \sim x^{3/2} \) as \( x \rightarrow \infty \) is consistent with the result in \eqref{AB} for
\[
\left|\frac{\nabla V}{V}\right| = 2\sqrt{\frac{2}{3}},
\]
which corresponds to the falloff rate for a scalar potential sourced by a higher-dimensional cosmological constant.

\section{Scale Separation and the Decoupling Test}\label{test}

In this section, we combine the results of Sections~\ref{sec2},~\ref{dec}, and~\ref{DC} to show that AdS spaces that are scale-separated, or more precisely those that violate the Asymptotic No-Scale Separation condition proposed in Section~\ref{ANSS}, cannot be realized as the near-horizon geometry of an extremal black brane that can be decoupled from the ambient spacetime gravity. In other words, we demonstrate that the CFT dual to such an AdS space cannot arise from decoupling the worldvolume theory of a brane. Consequently, if such an AdS space exists, its CFT dual cannot be embedded in string theory. We now provide a summary of the argument, accompanied by a brief review of the relevant results established in the preceding sections.

In standard holography, one takes a decoupling limit of a brane to argue that its worldvolume theory becomes dual to the near-horizon AdS geometry. This limit focuses on the low-energy modes localized on the brane or trapped in the throat, such that their coupling to the asymptotic gravitational sector becomes negligible. In Section~\ref{dec}, we showed that the decoupling limit exists only if a specific asymptotic condition is satisfied, which we called the \textit{decoupling condition}. This condition requires
\begin{align}\label{ratio}
    \lim_{\rho \rightarrow \infty} \frac{M_{\rm pl}}{f(\rho)\, \Lambda_s(\rho)}<\infty
\end{align}
where the extremal black brane background takes the form
\begin{align}\label{nte1}
ds^2 = f(\rho)^2\left[-dt^2 + d\vec{y}_p^2\right] + d\rho^2 + \cdots,
\end{align}
with the ellipsis denoting additional transverse directions.

To test this condition, in Section~\ref{uplift} we developed a general method for constructing black brane geometries from any AdS theory. Briefly, a black brane corresponds to a scalar field solution with two spatial infinities. At one end, the scalar field stabilizes at a critical point of the potential, giving rise to an AdS throat. At the other end, the scalar field runs off to infinity, corresponding to the asymptotic region far from the brane. These solutions can be described by scalar field theory coupled to gravity in a lower-dimensional effective theory obtained by dimensional reduction over the transverse directions. The reduced metric takes the form
\begin{align}\label{te2}
ds^2 = a(x)^2\left[-dt^2 + d\vec{y}_p^2\right] + dx^2.
\end{align}

Since the scalar field diverges as \( x \rightarrow \infty \), the asymptotic behavior of the scalar potential determines whether the extremal black brane can admit a decoupling limit. In particular, we show that the existence of a decoupling limit implies that the scalar potential must satisfy the Asymptotic No-Scale Separation condition
\[
\frac{V}{V}\cdot\frac{\nabla\Lambda_s}{\Lambda_s}\leq\frac{2}{d-2}\,,
\]
which as we explained in Section~\ref{ANSS} implies that the scalar potential cannot decay faster than \( m^{2} \) in the direction of steepest descent of $\Lambda_s$. Therefore, we conclude that the existence of a decoupling limit implies that the Asymptotic No-Scale Separation condition must be satisfied. 

For AdS spaces that can be constructed in string theory, we show that if a given construction violates the Asymptotic No-Scale Separation condition, then the CFT dual to that AdS cannot be realized as a worldvolume theory of a brane in string theory. This is a highly non-trivial result, since the completeness of string theory for SCFTs has been extensively and successfully used to classify conformal field theories across different dimensions and with varying amounts of supersymmetry.

\subsection{The Asymptotic No Scale Separation Condition}\label{MA}

Let us now proceed with the detailed argument. First, we show how the decoupling test expressed in terms of the black brane background \eqref{nte1} can also be evaluated in the dimensionally reduced background \eqref{te2}, obtained by truncating on the non-radial transverse directions.

Note that the functions \( a(x) \) and \( f(\rho) \) in \eqref{te2} and \eqref{nte1}, respectively, are related via a Weyl rescaling that maps the induced metric in the \( (t, \vec{y}_p, \rho) \) subspace to the lower-dimensional Einstein-frame metric in coordinates \( (t, \vec{y}_p, x) \).
This rescaling introduces a spacetime-dependent factor \( C \), such that
\begin{align}
    f^2 = C\, a^2.
\end{align}
Under this Weyl rescaling, the higher-dimensional Einstein-Hilbert term transforms as
\begin{align}
    \left(\sqrt{g}\, \mathcal{R}\right)_{\rm higher~dim.} \rightarrow C^{\frac{d-2}{2}} \left(\sqrt{g}\, \mathcal{R}\right)_{\rm lower~dim.}.
\end{align}
The prefactor \( C^{\frac{d-2}{2}} \) is absorbed into a redefinition of the Planck scale, leading to the relation
\begin{align}
    \left( \frac{M_{\rm pl,D}}{\Lambda_s} \right)^{d-2} = \left( \frac{M_{\rm pl}}{\Lambda_s} \right)^{d-2} C^{\frac{d-2}{2}},
\end{align}
where \( M_{\rm pl,D} \) is the Planck scale in \( D \)-dimensions, and \( M_{\rm pl} \) is the lower-dimensional Planck scale. Combining this with the relation between \( f \) and \( a \), we conclude that the dimensionless decoupling ratio transforms as
\begin{align}\label{WRC}
    f(\rho)^{-1}\left(\frac{M_{\rm pl,D}}{\, \Lambda_s}\right)(\rho) = a(x)^{-1}\left(\frac{M_{\rm pl}}{\, \Lambda_s}\right)(x).
\end{align}
Note that we work with the dimensionless ratios as position-dependent functions in order to avoid any ambiguity regarding units and frames. For example, the higher-dimensional and the lower-dimensional Einstein frames fix their respective Planck scales to be constant and express all quantities in those units. Based on~\eqref{WRC}, the decoupling condition can be tested directly in the lower-dimensional theory by checking whether
\begin{align}
    g_{\rm grav.}=\lim_{x \rightarrow \infty} \frac{M_{\rm pl}}{a(x)\, \Lambda_s(x)}
\end{align}
remains finite.

We now show that it is impossible to find a scalar potential with an AdS critical point that satisfies the condition
\begin{align}\label{cond}
    \nabla \ln(V)\cdot\nabla\ln(\Lambda_s)\geq 2 + \epsilon
\end{align}
in all infinite-distance limits along the directions of $\nabla V$, for some positive \( \epsilon \). We will assume that such a potential exists and prove the above statement by contradiction.

Let us consider an infinite-distance limit in the steepest-descent direction of the potential, which is associated with a scalar field \( \phi \). Suppose in this direction,
\[
V(\phi) \sim e^{-\lambda \phi}, \quad \Lambda_s(\phi) \sim e^{-c \phi}.
\]
Equation \eqref{cond} implies that 
\begin{align}\label{cbound}
    c\geq \frac{2+\epsilon}{(d-2)\lambda}\,.
\end{align}
Depending on the value of \( \lambda \), we characterized the behavior of \( a(x) \) at large \( x \) in equation \eqref{AB}. We summarize the results here:

\begin{itemize}
    \item \textbf{Case I:} \( \lambda > 2\sqrt{\frac{d-1}{d-2}} \)
    \begin{align*}
        a(x) &\sim x^{\frac{1}{d-1}}, \\
        \phi(x) &\simeq \sqrt{\frac{d-2}{d-1}} \ln(x).
    \end{align*}
    
    \item \textbf{Case II:} \( \lambda \leq 2\sqrt{\frac{d-1}{d-2}} \)
    \begin{align*}
        a(x) &\sim x^{\tfrac{4}{(d-2)\lambda^2}}, \\
        \phi(x) &\simeq \frac{2}{\lambda} \ln(x).
    \end{align*}
\end{itemize}

\noindent Let us now consider each case and test the decoupling condition:
\begin{align}
    \lim_{x \rightarrow \infty} \frac{M_{\rm pl}}{a(x)\, \Lambda_s(x)} < \infty\,.
\end{align}

\paragraph{Case I:}
\[
\frac{M_{\rm pl}}{a(x)} \sim x^{-\frac{1}{d-1}} \sim e^{-\frac{\phi}{\sqrt{(d-1)(d-2)}}}.
\]
From \eqref{ESCA}, we know that \( \Lambda_s \) decreases at least as fast as the RHS, with saturation only occurring in the limit corresponding to decompactification to one higher dimension. Thus,
\[
g_{\rm grav.} = \frac{M_{\rm pl}}{a(x)\Lambda_s(x)} \to \infty,
\]
unless exactly one dimension decompactifies far from the brane. However, such a setup implies a codimension-2 brane, for which a large \( N \) limit is inconsistent due to the emergence of a large conical deficit and eventual overclosure.

\paragraph{Case II:}
\[
\frac{M_{\rm pl}}{a(x)} \sim x^{-\tfrac{4}{(d-2)\lambda^2}}\sim e^{-\frac{2\phi}{(d-2)\lambda}}.
\]
From \eqref{cbound}, we get \( c \geq \frac{2+\epsilon}{(d-2)\lambda} \), and hence:
\[
g_{\rm grav.} = \frac{M_{\rm pl}}{a(x)\Lambda_s(x)} \sim e^{(c - \frac{2}{(d-2)\lambda}) \phi} \to \infty.
\]

In both cases, the decoupling test fails. We therefore conclude that our assumption that the Asymptotic No Scale Separation Condition could be violated was incorrect, completing the proof by contradiction.

Let us now make an important comment. The argument presented here essentially shows that unless the scalar potential satisfies the ANSS condition, the energies of perturbative modes will exceed the species scale, leading to a breakdown of the effective field theory. This breakdown corresponds precisely to the absence of a decoupling limit for modes at spacetime infinity. In fact, this is the spacelike analog of the argument presented in~\cite{Bedroya:2025ris} for cosmologies realized in the string landscape.

In that work, the authors (two of whom are authors of the present paper) demonstrated that in flat FRW cosmologies driven by scalar potentials, unless the potential vanishes at infinity more slowly than, or at the same rate as, \( m^2 \) (where \( m \) denotes the mass of an infinite tower of states), the energy of a generic perturbation will exceed the quantum gravity cutoff, again leading to a breakdown of EFT. It was emphasized that this is a genuine failure of effective field theory, unlike the so-called "trans-Planckian problem" in cosmology, which—though conjectured to be problematic in quantum gravity~\cite{Bedroya:2019snp}, is not excluded by EFT alone. Moreover, in the cosmological context, it is necessary for the scalar potential to remain below \( m^2 \) to ensure that the Hubble scale remains below the tower scale, thereby maintaining the validity of the lower-dimensional EFT~\cite{Rudelius:2022gbz}.

As discussed in Section~\ref{Wick}, the double Wick rotation provides a one-to-one map between these cosmological arguments and their black brane analogs. However, this analogy has limitations in the black brane context. In particular, we are not necessarily concerned with whether perturbations in the black brane background are fully captured by a lower-dimensional EFT describing the brane. An example where this fails is given by AdS spaces embedded in higher-dimensional AdS spacetimes, reviewed in Section~\ref{AdSdS}. As we explained there, some limits allow the scalar potential to scale with the square of the higher-dimensional Planck scale, thereby violating the \( |V| \lesssim m^2 \) condition.

In~\cite{Bedroya:2025ris}, it was shown that the breakdown of EFT at future infinity in cosmology can be avoided by considering spatially curved FRW spacetimes with negative curvature. In such spacetimes, the late-time behavior of perturbations is modified so that their energies no longer exceed the quantum gravity cutoff. By contrast, it was also shown in~\cite{Bedroya:2025ris} that positive spatial curvature is insufficient, as it inevitably leads to a future singularity. Thus, one must instead consider hyperbolic spatial slices.

One might wonder whether a similar resolution could be applied in the black brane context by considering alternative slicings. Translating this idea via Double Wick Rotation amounts to replacing the brane-parallel Minkowski geometry with de Sitter slices:
\begin{align}
ds^2 = f(\rho)^2 ds^2_{dS_{p+1}} + d\rho^2 + \cdots\,,
\end{align}
where $ds^2_{dS_{p+1}}$ denotes the $(p+1)$-dimensional de Sitter metric with unit Hubble radius. Such solutions, however, do not develop an infinite throat and are therefore unsuitable for holography. To see this, note that if the scalar field stabilizes at a critical value, the geometry becomes AdS, for which the warp factor takes the form $f(\rho) = L_{\rm AdS}\sinh(\rho/L_{\rm AdS})$. This coordinate system terminates at a finite value (here $\rho=0$), leaving no infinite throat. Although these geometries are mathematically consistent, they do not admit the infinite-throat structure required for holographic duality.

\subsection{Implications for DGKT and KKLT}\label{scalesep}

Let us now comment on the implications and limitations of our results for scale separation. Parametric scale separation means that one can find a family of AdS vacua for which the ratio $\frac{V_{\rm AdS}}{m^2}$ can be made arbitrarily small in Planck units, where \( V_{\rm AdS} \) is the value of the scalar potential at the critical point corresponding to the AdS space. 

The reason \( V_{\rm AdS} \) and \( m^2 \) are typically related is that the scalar potential usually receives a contribution sourced by the curvature of the internal space, scaling with \( m^2 \). Since different contributions compete at the critical point, the critical value of the potential generically ends up being of order \( m^2 \). Importantly, this curvature-sourced contribution also persists in the asymptotic regions of field space. Therefore, the lack of scale separation is natural for theories in which the scalar potential scales as \( m^2 \) in some infinite-distance limit, since the mass scale of the corresponding tower in that direction will be of the same order as \( |\Lambda_{\rm AdS}|^{1/2} \). This need not be the case for all infinite-distance limits, as there may exist towers much heavier than \( |\Lambda_{\rm AdS}|^{1/2} \) which become light in other directions.

In Section~\ref{attempts}, we discussed two natural approaches to constructing a family of scale-separated vacua. The two strategies are: 
\begin{enumerate}
    \item To find a family of compactifications that have an \( m^2 \)-contribution to the scalar potential, but where the proportionality constant $V/m^2$ can be made arbitrarily small. For example, one might achieve this by performing a Freund--Rubin compactification on a family of Einstein manifolds with a vanishing ratio of Ricci scalar to \( m_{\rm KK}^2 \).
    \item To instead consider compactifications in which the scalar potential vanishes faster than \( m^2 \) in all weak coupling limits, i.e. $\lim_{\phi\rightarrow\infty}\frac{m^2}{|V|}\rightarrow\infty$. 
\end{enumerate}
We restrict our attention to the second class, since as explained in Section~\ref{attempts}, there is compelling mathematical evidence that the first class cannot be realized~\cite{Collins:2022nux}. However, it is important to emphasize that a formal general proof is still lacking.
The second class, by contrast, includes compactifications with flat internal spaces where fluxes dominate the potential. Examples include the DGKT and KKLT scenarios, in which massive type IIA or type IIB supergravity are compactified on Calabi--Yau orientifolds, respectively. We reviewed the DGKT scenario in more detail in Section~\ref{attempts}, noting that it is arguably the best-established scale-separated AdS construction. A well-studied example of DGKT is the compactification of massive type IIA on a $T^6/(\mathbb{Z}_3 \times \mathbb{Z}_3)$ orientifold. In this case, the scalar potential stabilizes the dilaton in terms of the $KK$ mass of the torus for large $T^6$ volume, such that 
\[
g_s \sim \left( \frac{l_{\rm KK}}{l_{\rm pl}} \right)^{-3/7},
\]
and the scalar potential in terms of the $KK$ mass takes the form 
\begin{align}
V \simeq M_{\rm pl}^4 \left[
A N^2 \left( \frac{m_{\rm KK}}{M_{\rm pl}} \right)^{26/7}
- B \left( \frac{m_{\rm KK}}{M_{\rm pl}} \right)^{18/7}
\right],
\end{align}
where $A$ and $B$ are constants. The first term is sourced by the 4-form fluxes, while the second term arises from balancing the contribution of the Romans mass with the orientifold tension. As is evident from the scaling $V \sim m_{\rm KK}^{18/7}$, the AdS critical point can be parametrically scale-separated in the large-$N$ limit. More importantly, in this limit both the string coupling and the compactification volume go to zero.

Even though string perturbation theory is under control in the truncated four-dimensional equations of motion, as explained in Section~\ref{AdSdist}, the AdS theory at finite $N$ cannot be arbitrarily weakly coupled. This is because any given AdS space only has a finite region of moduli space where the gravitational (supergravity) description is valid. Once couplings are taken to extreme values, the CFT dual becomes the appropriate description. Therefore, the supergravity approximation cannot be precisely defined even in the weak-coupling regime and always breaks down beyond a finite patch of moduli space. Unlike Minkowski quantum gravity, where one can often take the coupling to zero, the UV-complete understanding of AdS is necessarily provided by its CFT dual.\footnote{There are examples such as M-theory where there is no coupling that can be taken to zero. However, in those cases, the evidence for the existence of 11-dimensional supergravity as a UV-complete theory comes from its emergence via a web of dualities, in particular as the strong-coupling limit of type IIA string theory.} 
Now, applying the argument of Subsection~\ref{MA} to the DGKT scenario, we find that the black brane realizing DGKT as its near-horizon geometry is problematic, since ten-dimensional gravity far from the brane becomes infinitely strongly coupled. In other words, although one might naively expect that string perturbation theory is arbitrarily reliable far from the brane—because the string coupling goes to zero, the energy of any given mode, measured in string units, actually diverges. While this is a special case of the general argument from the previous subsection, it is nevertheless instructive to repeat the derivation carefully here.

Suppose $\hat\phi$ and $\hat\rho$ are respectively the canonically normalized 10D dilaton and the volume modulus. Then the string length and KK length in units of the 4D Planck scale are given by
\begin{align}
    \frac{l_s}{l_{\rm pl}} &\propto \exp\!\left(\tfrac{1}{\sqrt{8}} \hat\phi + \sqrt{\tfrac{3}{8}} \hat\rho\right), \\
    \frac{l_{\rm KK}}{l_{\rm pl}} &\propto \exp\!\left(\sqrt{\tfrac{2}{3}} \hat\rho\right).
\end{align}
In the gradient direction of the scalar potential in the large-volume limit we have 
\[
l_s \, l_{\rm KK}^{-6/7} \propto l_{\rm pl}^{1/7}.
\]
Thus, the gradient direction of $V$ is one in which the combination
\[
\frac{1}{\sqrt{8}}\hat\phi - \left(\frac{2\sqrt{6}}{7} - \sqrt{\tfrac{3}{8}}\right)\hat\rho
\]
is kept fixed. Let us define $\varphi$ to be the canonically normalized distance in this direction. Then
\begin{align}
    \frac{\partial \hat\rho}{\partial \varphi} = 
    \frac{1/\sqrt{8}}{\sqrt{ \tfrac{1}{8} + \left( \frac{2\sqrt{6}}{7} - \sqrt{\tfrac{3}{8}} \right)^2 }}=\frac{7\sqrt{13}}{26}.
\end{align}
Therefore, in this direction we have
\begin{align}
    V \propto \left(\frac{m_{\rm KK}}{M_{\rm pl}} \right)^{18/7} M_{\rm pl}^4 
      \propto \exp\!\left(-\tfrac{3\sqrt{6}}{\sqrt{13}} \varphi\right).
\end{align}
Since the coefficient $\tfrac{3\sqrt{6}}{\sqrt{13}}$ is smaller than $\sqrt{6}$, according to \eqref{AB}, we find that the black brane solution is a four-dimensional solution 
\begin{align}
ds^2 = a(x)^2\left[-dt^2 + d\vec{y}_p^2\right] + dx^2,
\end{align}
such that
\begin{align}
        x \to -\infty &: \quad a(x) \sim e^{x/L_{\rm AdS}}, 
        && \varphi(x) \to \text{constant}, \\
        x \to \infty &: \quad a(x) \sim x^{13/27}, 
        && \varphi(x) \simeq \sqrt{\tfrac{26}{27}} \ln(x).
\end{align}

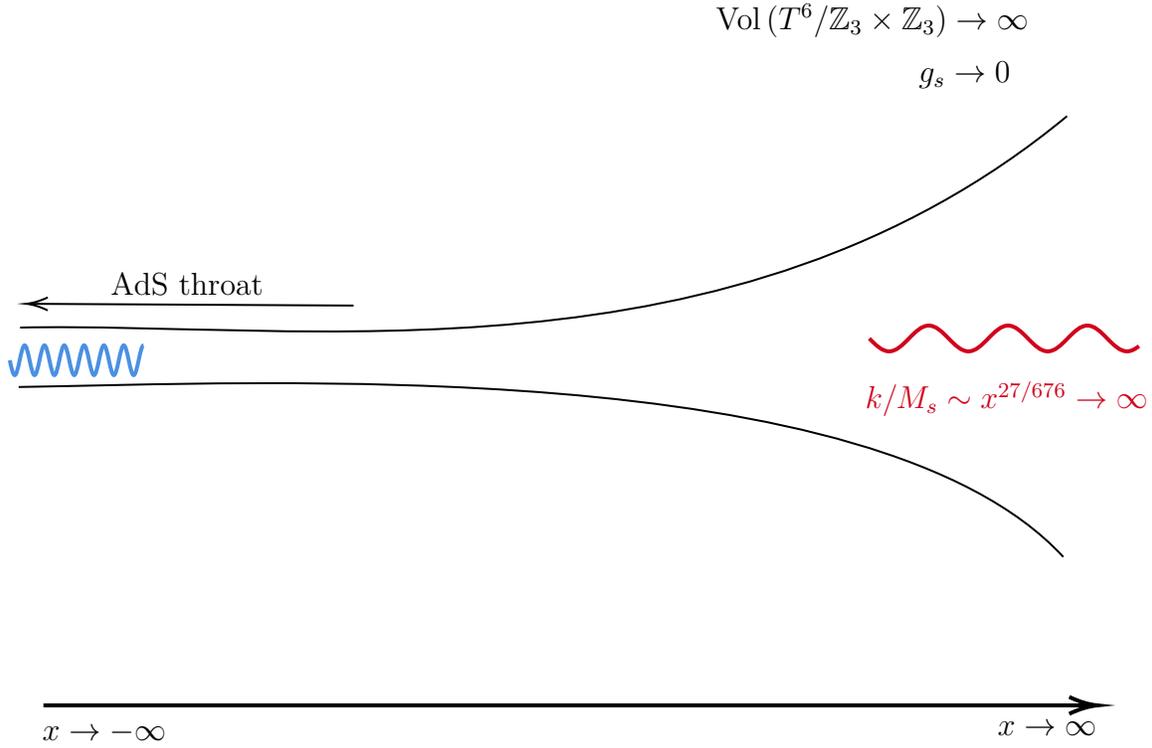
\begin{figure}
    \centering

  
\tikzset {_1dit9667t/.code = {\pgfsetadditionalshadetransform{ \pgftransformshift{\pgfpoint{0 bp } { 0 bp }  }  \pgftransformrotate{0 }  \pgftransformscale{2 }  }}}
\pgfdeclarehorizontalshading{_2y565rftr}{150bp}{rgb(0bp)=(1,1,0);
rgb(37.5bp)=(1,1,0);
rgb(62.5bp)=(0,0.5,0.5);
rgb(100bp)=(0,0.5,0.5)}
\tikzset{every picture/.style={line width=0.75pt}} 

\begin{tikzpicture}[x=0.75pt,y=0.75pt,yscale=-1,xscale=1]

\draw [shading=_2y565rftr,_1dit9667t][line width=1.5]    (51.97,378.7) -- (580.52,378.7) ;
\draw [shift={(583.52,378.7)}, rotate = 180] [color={rgb, 255:red, 0; green, 0; blue, 0 }  ][line width=1.5]    (14.21,-4.28) .. controls (9.04,-1.82) and (4.3,-0.39) .. (0,0) .. controls (4.3,0.39) and (9.04,1.82) .. (14.21,4.28)   ;
\draw    (40.24,188) .. controls (181.73,184.94) and (399.78,220.68) .. (568.22,81.29) ;
\draw    (39.49,218) .. controls (157.87,216.21) and (470.9,200.13) .. (566.35,303.77) ;
\draw    (208.4,176.9) -- (45.23,176.01) ;
\draw [shift={(43.23,176)}, rotate = 0.31] [color={rgb, 255:red, 0; green, 0; blue, 0 }  ][line width=0.75]    (10.93,-3.29) .. controls (6.95,-1.4) and (3.31,-0.3) .. (0,0) .. controls (3.31,0.3) and (6.95,1.4) .. (10.93,3.29)   ;
\draw  [color={rgb, 255:red, 74; green, 144; blue, 226 }  ,draw opacity=1 ][line width=1.5]  (35,204.5) .. controls (35.82,208.34) and (36.6,212) .. (37.5,212) .. controls (38.4,212) and (39.18,208.34) .. (40,204.5) .. controls (40.82,200.66) and (41.6,197) .. (42.5,197) .. controls (43.4,197) and (44.18,200.66) .. (45,204.5) .. controls (45.82,208.34) and (46.6,212) .. (47.5,212) .. controls (48.4,212) and (49.18,208.34) .. (50,204.5) .. controls (50.82,200.66) and (51.6,197) .. (52.5,197) .. controls (53.4,197) and (54.18,200.66) .. (55,204.5) .. controls (55.82,208.34) and (56.6,212) .. (57.5,212) .. controls (58.4,212) and (59.18,208.34) .. (60,204.5) .. controls (60.82,200.66) and (61.6,197) .. (62.5,197) .. controls (63.4,197) and (64.18,200.66) .. (65,204.5) .. controls (65.82,208.34) and (66.6,212) .. (67.5,212) .. controls (68.4,212) and (69.18,208.34) .. (70,204.5) .. controls (70.82,200.66) and (71.6,197) .. (72.5,197) .. controls (73.4,197) and (74.18,200.66) .. (75,204.5) .. controls (75.82,208.34) and (76.6,212) .. (77.5,212) .. controls (78.4,212) and (79.18,208.34) .. (80,204.5) .. controls (80.82,200.66) and (81.6,197) .. (82.5,197) .. controls (83.4,197) and (84.18,200.66) .. (85,204.5) .. controls (85.82,208.34) and (86.6,212) .. (87.5,212) .. controls (88.4,212) and (89.18,208.34) .. (90,204.5) .. controls (90.82,200.66) and (91.6,197) .. (92.5,197) .. controls (93.4,197) and (94.18,200.66) .. (95,204.5) .. controls (95.82,208.34) and (96.6,212) .. (97.5,212) .. controls (98.4,212) and (99.18,208.34) .. (100,204.5) .. controls (100.82,200.66) and (101.6,197) .. (102.5,197) ;
\draw  [color={rgb, 255:red, 208; green, 2; blue, 27 }  ,draw opacity=1 ][line width=1.5]  (468.5,193.5) .. controls (471.76,196.83) and (474.88,200) .. (478.5,200) .. controls (482.12,200) and (485.24,196.83) .. (488.5,193.5) .. controls (491.76,190.17) and (494.88,187) .. (498.5,187) .. controls (502.12,187) and (505.24,190.17) .. (508.5,193.5) .. controls (511.76,196.83) and (514.88,200) .. (518.5,200) .. controls (522.12,200) and (525.24,196.83) .. (528.5,193.5) .. controls (531.76,190.17) and (534.88,187) .. (538.5,187) .. controls (542.12,187) and (545.24,190.17) .. (548.5,193.5) .. controls (551.76,196.83) and (554.88,200) .. (558.5,200) .. controls (562.12,200) and (565.24,196.83) .. (568.5,193.5) .. controls (571.76,190.17) and (574.88,187) .. (578.5,187) .. controls (582.12,187) and (585.24,190.17) .. (588.5,193.5) .. controls (591.76,196.83) and (594.88,200) .. (598.5,200) .. controls (600.63,200) and (602.59,198.9) .. (604.5,197.32) ;

\draw (389.82,23.2) node [anchor=north west][inner sep=0.75pt]    {$\text{Vol}\left( T^6/\mathbb{Z}_3 \times \mathbb{Z}_3\right)\rightarrow \infty $};
\draw (531.85,385.4) node [anchor=north west][inner sep=0.75pt]    {$x\rightarrow \infty $};
\draw (50.05,385.4) node [anchor=north west][inner sep=0.75pt]    {$x\rightarrow -\infty $};
\draw (84.4,159.01) node [anchor=north west][inner sep=0.75pt]   [align=left] {AdS throat};
\draw (492,52.4) node [anchor=north west][inner sep=0.75pt]    {$g_{s}\rightarrow 0$};
\draw (465,213.4) node [anchor=north west][inner sep=0.75pt]  [color={rgb, 255:red, 208; green, 2; blue, 27 }  ,opacity=1 ]  {$k/M_{s} \sim x^{27/676}\rightarrow \infty $};

\end{tikzpicture}
\caption{The figure illustrates the black brane solution that realizes the DGKT AdS vacuum as its near-horizon geometry. In standard holography, the worldvolume theory on the brane in the regime where the gravitational description breaks down is dual to the DGKT AdS. However, in this case, even though far from the brane the string coupling vanishes and the volume of $T^6$ diverges, the energy of any given mode in units of the string scale diverges. Note that the energy of each mode vanishes in Planck units at infinity. It is precisely because of the decreasing string coupling that the mode energies diverge in string units as $x \to \infty$, rendering such modes impossible to capture within the string worldsheet description.}
    \label{DGKTbrane}
\end{figure}

Far away from the brane, we then find
\begin{align}
    \frac{m_s}{m_{\rm pl}} = \left(\frac{l_s}{l_{\rm pl}}\right)^{-1} 
    \propto \exp\!\left(-\tfrac{1}{\sqrt{8}}\hat\phi - \sqrt{\tfrac{3}{8}}\hat\rho\right) 
    = \exp\!\left(-\sqrt{\tfrac{6}{13}} \varphi\right) 
    \propto x^{-\sqrt{12/27}}.
\end{align}
Since $\tfrac{m_s}{m_{\rm pl}}$ decreases faster than $a^{-1} \sim x^{13/27}$, we conclude that the momentum of any mode in string units,
\[
\frac{k}{m_s}(x) \propto a^{-1}\,\frac{m_{\rm pl}}{m_s},
\]
diverges as $x \to \infty$. On the worldsheet, the ratio $k/m_s$ enters the conformal weight of the corresponding vertex operator, 
\[
\Delta(k) = \Delta_0 + \tfrac{1}{2}\!\left(\tfrac{k}{m_s}\right)^2.
\]
Thus, any mode corresponds to a vertex with infinite conformal weight and cannot be captured by string perturbation theory, despite the fact that the string coupling vanishes and the internal volume diverges (Figure~\ref{DGKTbrane}). This means that string theory cannot be consistently formulated in this background.

We therefore conclude that the holographic dual of DGKT cannot be realized as a brane in string theory, since the associated worldvolume theory does not admit a decoupling limit. Note that the obstruction is not the mere inclusion of orientifolds, but rather their presence in a compact internal space. Placing orientifolds in a compact manifold necessitates turning on the Romans mass to cancel the $F_2$ tadpole. However, the Romans mass in turn generates a scalar potential for the dilaton, which drives the energy of any given mode far from the brane to exceed the string scale. In this sense, the compactness of the internal manifold is the crucial ingredient which is required in order to obtain a lower-dimensional theory of gravity.

Let us also comment on the KKLT scenario~\cite{Kachru:2003aw}. The KKLT scenario is a flux compactification of type IIB string theory on an orientifold Calabi--Yau threefold such that, to leading order, the internal manifold has no curvature. A scale-separated AdS is then constructed by balancing the flux superpotential with non-perturbative corrections, and the vacuum is uplifted by introducing anti-D3 branes. Constructing a scale-separated AdS is thus an essential intermediate step in the KKLT framework.

To test the KKLT scenario, one must analyze the infinite-distance limits along the gradient directions of the scalar potential and verify whether they satisfy the Asymptotic No-Scale-Separation condition. The moduli space of a generic KKLT flux compactification is more complex than that of the explicit DGKT example we considered.\footnote{Even before introducing fluxes, interesting infinite-distance limits appear in the $\mathcal{N}=2$ moduli space obtained by compactifying type II string theories on a Calabi--Yau threefold. The infinite-distance limits in the vector-multiplet moduli space are classified (see~\cite{Friedrich:2025gvs,Monnee:2025ynn} and references therein), whereas those in the hypermultiplet moduli space remain unclassified.} 
Nevertheless, because the internal manifold is Ricci-flat, one still expects a violation of the Asymptotic No-Scale-Separation condition which would imply $m^2\lesssim |V|$ along the direction of $-\nabla \Lambda_s$.

Following the earlier version of this work, it was shown in~\cite{Bedroya:2025fie} that the Asymptotic No-Scale-Separation condition is never satisfied in the KKLT scenario. The key insight is that for the dual brane to decouple, the scalar potential along the direction of $\nabla V$ must be negative and exponentially decaying. However, in KKLT compactifications, the tree-level potential is never negative, and when the non-perturbative contribution dominates, it does not exhibit an exponential falloff. Consequently, the condition fails to hold in all relevant infinite-distance directions.

Our test for the holographic consistency of KKLT is distinct from the one in~\cite{Lust:2022lfc}, where the authors analyzed the candidate dual of KKLT and argued that the KKLT AdS cannot arise in a regime where the gravitational description is under control. By contrast, we employ the asymptotic behavior of the scalar potential to argue for the lack of scale separation. As explained earlier, no supergravity construction of an AdS vacuum can establish UV-completeness by itself, since it lacks a weak-coupling limit. Therefore, the only way to ensure the UV-completeness of an AdS construction is to identify its CFT dual.

\section{Comments on dS/CFT}\label{dSR}

In this section, we apply the machinery developed in the rest of this paper to test the dS/CFT correspondence. In the previous sections, we focused on whether the CFT dual to a given AdS space can be realized as the worldvolume theory of a brane in string theory. Here, we ask an analogous question in the context of dS/CFT~\cite{Strominger:2001pn}. Specifically, suppose a de Sitter space (dS) can be realized in string theory, and assume that there exists a Euclidean CFT dual to it. Can such a CFT be realized as the worldvolume theory of a (spacelike) brane in string theory? 

We show that, unlike the AdS case where only scale-separated AdS vacua are problematic, the situation in de Sitter space is more restrictive: the dual CFT can never be realized as the worldvolume theory of a brane. Importantly, the conclusion of this section is no longer tied to scale separation. In fact, it is known that parametric scale separation can be achieved in cosmologies realized in string theory~\cite{Andriot:2025cyi,Bedroya:2025ris}.

Realizing dS space within string theory has proven to be challenging. It is widely understood that an eternally stable de Sitter space cannot be realized in string theory due to the behavior of scalar potentials: in any direction where the theory becomes weakly coupled, the potential energy tends to vanish~\cite{Dine:1985he}. As a result, any critical point of the scalar potential with positive energy cannot correspond to a global minimum. Consequently, any de Sitter vacuum is necessarily unstable, either perturbatively due to tachyonic directions, or non-perturbatively due to vacuum decay via bubble nucleation.

The first class of solutions is referred to as \textit{unstable de Sitter}, while the second corresponds to \textit{metastable de Sitter}. Unstable de Sitter constructions have been realized, for example, using duality arguments~\cite{Chen:2025rkb} or by combining Casimir energy and fluxes~\cite{ValeixoBento:2025yhz}. However, constructing metastable de Sitter vacua is considerably more difficult, as it requires a delicate balancing of multiple ingredients, each of which must remain under control. Two well-known proposals to construct metastable dS vacua along these lines are the KKLT scenario~\cite{Kachru:2003aw} and the Large Volume Scenario (LVS)~\cite{Balasubramanian:2005zx}. Yet the validity of these constructions have been subject to significant scrutiny in recent years~\cite{Moritz:2017xto,Gautason:2018gln,Hamada:2018qef,Hamada:2019ack,Carta:2019rhx,Gautason:2019jwq,Bena:2019mte,Kachru:2019dvo,Randall:2019ent,Gao:2020xqh,Bena:2020xrh,Demirtas:2021nlu,Junghans:2022exo,Gao:2022fdi,Lust:2022lfc,Lust:2022xoq,Hebecker:2022zme,Bena:2022ive,ValeixoBento:2023nbv,McAllister:2024lnt,Kim:2024dnw,Bena:2024are,Moritz:2025bsi}.

One potential approach to defining de Sitter quantum gravity without a direct string theory construction involves holography. Specifically, inspired by the AdS/CFT correspondence, a proposal known as the dS/CFT correspondence was introduced. In this framework, the bulk de Sitter geometry is conjectured to be holographically dual to the worldvolume theory of a spacelike brane, known as an S-brane.

The central idea of dS/CFT is that one can define boundary correlation functions on the spacelike boundary of de Sitter space that exhibit the scaling behavior characteristic of a conformal field theory~\cite{Strominger:2001pn}. These boundary correlators are obtained by taking $n$-point functions in the bulk, multiplying them by the inverse of the bulk-to-boundary propagators $G$, and taking the limit as the bulk points $x, y$ approach boundary points $X, Y$ at future infinity \( \mathscr{I}^+ \). More precisely, for a scalar field, one finds
\begin{align}
    \expval{\mathcal{O}_\pm(X)\mathcal{O}_\pm(Y)} 
    = \lim_{x \to X,\, y \to Y \atop x,y \to \mathscr{I}^+}
    \expval{\varphi_\pm(x)\varphi_\pm(y)}\, G^{-1}(x,X)\, G^{-1}(y,Y),
\end{align}
where the $\pm$ labels denote the two independent fall-off modes of the scalar field near the boundary. The bulk-to-boundary propagator \( G(x, X) \) encodes how a source at the boundary influences the field in the bulk, and its inverse removes this dependence to isolate the intrinsic boundary dependence.

If the mass of the scalar field is \( m \), the scaling behavior of the boundary two-point function is governed by conformal weights \( \Delta_\pm \), leading to
\begin{align}
    \expval{\mathcal{O}_\pm(X)\mathcal{O}_\pm(Y)} \sim \frac{1}{|\vec{X} - \vec{Y}|^{\Delta_\pm}},
\end{align}
with
\begin{align}
    \Delta_\pm = \frac{d-1}{2} \pm \sqrt{\left(\frac{d-1}{2}\right)^2 - m^2 L^2},
\end{align}
where \( L \) is the de Sitter Hubble radius and \( d \) is the spacetime dimension of the boundary. 

In Section~\ref{dec}, we showed that for the holographic dual of an AdS geometry to be interpreted as the worldvolume theory of a brane embedded in string theory, the corresponding brane must satisfy a decoupling condition. In Section~\ref{test} we saw that this requirement imposes non-trivial consistency constraints. In the present section, we examine the analogous condition for realizing the S-brane dual of de Sitter space in string theory.

Since the S-brane is spacelike, it must reside at a constant time slice, and the analogue of the ``near-horizon geometry'' corresponds to evolving toward the S-brane in the time direction. In fact, we have already encountered such solutions in the context of cosmological spacetimes that arise as double Wick rotations of black branes, as discussed in Section~\ref{Wick}.

Suppose we consider a scalar potential with a local positive maximum at \( \varphi = \varphi_0 \) that asymptotically vanishes, as illustrated in Figure~\ref{dSt}. For such a potential, one can construct an FRW cosmological solution in which, in the asymptotic past, the geometry develops a de Sitter throat, while in the far future the scalar field rolls off to infinity:
\begin{align}
    ds^2 &= -dt^2 + a(t)^2\, d\vec{x}^2, \nonumber\\
    t \to -\infty &: \quad a(t) \sim e^{t/L}, \quad \varphi \to \varphi_0, \nonumber\\
    t \to \infty &: \quad a(t) \to t^p, \quad \varphi \to \infty,
\end{align}
where \( p \) is determined by the asymptotic behavior of the scalar potential, as discussed in~\eqref{AB1} and~\eqref{AB2}:
\begin{align}
    \left|\frac{\nabla V}{V}\right| \geq 2\sqrt{\frac{d-1}{d-2}} \quad &\Rightarrow \quad p = \frac{1}{d-2}, \nonumber\\
    \left|\frac{\nabla V}{V}\right| \leq 2\sqrt{\frac{d-1}{d-2}} \quad &\Rightarrow \quad p = \frac{4}{(d-2)\left|\frac{\nabla V}{V}\right|^2}.
\end{align}

In this picture, the S-brane provides the holographic dual description of the de Sitter throat region. The limit \( t \to -\infty \) corresponds to the flat slicing of de Sitter space, analogous to the Poincaré patch of AdS often used in near-horizon black brane geometries.

However, the de Sitter region that appears in the throat in the infinite past is necessarily unstable. This is evident from Figure~\ref{dSt}: in the throat, all the energy is stored in the scalar potential. To smoothly connect this de Sitter region to the future asymptotics via a classical solution, the scalar must convert potential energy into kinetic energy. This requires the potential to have a tachyonic direction near its maximum.

Therefore, in order for the S-brane dual to de Sitter space to be realized in string theory—that is, to admit a decoupled brane description—the corresponding de Sitter geometry must be unstable. This implies the presence of a scalar field with \( m^2 < 0 \), which leads to a serious problem for dS/CFT. Depending on the scalar mass \( m \), the conformal weights \( \Delta_\pm \) can acquire imaginary components, indicating a non-unitary boundary theory. But a qualitatively worse behavior occurs when \( m^2 < 0 \), as then \( \Delta_- < 0 \), and the two-point function fails to decay at large separations:
\begin{align}
    |\vec{X} - \vec{Y}| \to \infty \quad \Rightarrow \quad \expval{\mathcal{O}_-(X)\mathcal{O}_-(Y)} \nrightarrow 0.
\end{align}
This violates cluster decomposition, and implies that the boundary theory is not a local CFT—not even a non-unitary one. Hence, for a dS/CFT correspondence to hold, even in the non-unitary case, the de Sitter phase must be metastable. However, if it is metastable, then the S-brane dual does not decouple and cannot arise from a controlled string theory construction. \textit{We thus conclude that no holographic de Sitter constructions exist within string theory.}

Let us also consider the possibility of more sophisticated slicings. In global coordinates, a comoving observer reaches a bounce in finite proper time. By tuning the kinetic energy of the scalar field near the bounce to be arbitrarily small, one can engineer a de Sitter throat of extended duration. While quantum fluctuations place a lower bound on the kinetic energy, an even deeper obstruction arises from the scalar potential's asymptotics. In global slicing, the spatial curvature is positive and as the scalar field goes to infinity, the spacetime could collapse. Specifically, if 
\begin{align}
    \left|\frac{\nabla V}{V}\right| \geq \frac{2}{\sqrt{d - 2}},
\end{align}
then the spacetime undergoes collapse. This inequality has been conjectured to universally hold in the infinite distance limits of quantum gravity~\cite{Bedroya:2022tbh,Rudelius:2021azq}. More importantly, there exist independent bottom-up derivations of this bound based on holographic principles~\cite{Bedroya:2022tbh} as well as from the Emergent String Conjecture~\cite{Bedroya:2025ris}, both of which use assumptions that hold in string theory. Thus, even in global coordinates, where one might hope to engineer a prolonged de Sitter throat by tuning the scalar field's kinetic energy near the bounce, the asymptotic behavior of the scalar potential rules out the existence of a consistent global de Sitter region as the near-horizon geometry of an S-brane.

\section{Conclusions}\label{Conc}

We have shown that achieving parametric scale separation in AdS is in tension with holography. To establish this, we first reframed AdS scale separation in terms of the behavior of the scalar potential at infinite-distance limits. The resulting condition, which we termed the \emph{Asymptotic No-Scale-Separation condition}, states that for any scalar potential with an AdS critical point, there must exist an infinite-distance limit along $\nabla V$ in which $\partial_\phi\ln(V)\partial_\phi\ln(\Lambda_s)\leq2/(d-2)$. Ricci-flat compactifications attempt to realize scale separation precisely by violating this condition. 

We then related this condition to holography by considering an AdS effective field theory violating it and constructing the black brane whose near-horizon geometry is the given AdS. We demonstrated that this near-horizon AdS geometry cannot be decoupled from the ambient spacetime gravity. Equivalently, in the regime where the brane theory should be described by a CFT, that CFT fails to decouple from gravity. Thus, the CFT dual to the AdS cannot be realized as the worldvolume theory of a brane, which by definition requires a decoupling regime in which the brane degrees of freedom decouple from the surrounding gravitational theory. As a concrete example, we showed that the brane dual to DGKT cannot be realized in string theory, which we interpret as evidence that DGKT does not exist.

Our results have important implications: for any proposed scale-separated AdS construction, one must explicitly verify that the Asymptotic No-Scale-Separation condition is satisfied in order to ensure that it admits a holographic dual, and hence a UV-complete embedding in string theory. After the earlier version of this paper appeared, this test was carried out for the KKLT scenario in~\cite{Bedroya:2025fie}, where it was shown that the condition is violated. It would be particularly interesting to apply this criterion to other proposed scale-separated AdS constructions.

In order to develop a systematic mechanism for identifying the black brane corresponding to a given AdS, we showed that there exists a mapping between warped stationary spacetimes of black branes and certain FRW cosmologies with the opposite sign of the scalar potential. This mapping, which we referred to as a \emph{double Wick rotation}, provides a powerful tool for studying black brane solutions. In particular, it shows that not every critical point of a scalar potential can be realized as the near-horizon geometry of a brane when other critical points intervene along the trajectory connecting the AdS critical point to the asymptotic region of field space. It would be very interesting to extract a precise condition characterizing when such black brane geometries exist. 

Finally, we extend our analysis to the dS/CFT correspondence. In contrast to AdS, where the obstruction arises only for scale-separated vacua, we find that in de Sitter space no dual CFT can be consistently interpreted as the worldvolume theory of a spacelike brane.

\section*{Acknowledgments}
We are indebted to Juan Maldacena, Miguel Montero, Irene Valenzuela, and Cumrun Vafa for valuable and helpful conversations during the completion of this paper. We are also grateful to Zihni Kaan Baykara,  Jos\'e Calder\'on-Infante, Igor Klebanov, Jacob McNamara, Georges Obied, Houri Tarazi, and Timo Weigand for valuable discussions. We are also grateful to Fien Apers, Thomas Eran Palti, Van Riet, and Timm Wrasse for providing feedback on this work. AB gratefully acknowledges the support and hospitality of the Simons Center for Geometry and Physics, Stony Brook University, during the 2025 Simons Physics Summer Workshop, where part of this work was carried out. AB is supported in part by the Simons Foundation 
under grant number~654561 and by the Princeton Gravity Initiative at 
Princeton University. PJS is supported in part by the U.S.\ Department  of Energy under grant number~DE-FG02-91ER40671 and by the Simons  Foundation under grant number~654561.

\appendix
\bibliographystyle{JHEP}
\bibliography{References.bib}

\end{document}